\documentclass[aps,prd,twocolumn,showpacs,superscriptaddress,nofootinbib]{revtex4}

\usepackage{amsmath}   
\usepackage{graphicx}  
\usepackage{xspace}
\usepackage{units}

\usepackage{hyperref}

\newcommand{\minerva}{MINERvA\xspace}

\newcommand{\nubar}{\ensuremath{\bar{\nu}}}

\newcommand{\sizecheck}{0} 
\newcommand{\PRLsupp}{1}   
\newcommand{\SuppLocation}{in the Appendix}

\newif\ifpdf
\ifx\pdfoutput\undefined
   \pdffalse
\else
   \pdfoutput=1
   \pdftrue
\fi
\ifpdf
   \usepackage{graphicx}
   \usepackage{epstopdf}
\else
   \usepackage{graphicx}
\fi

\begin{document}
\title{Measurement of Neutrino Flux from Neutrino-Electron Elastic Scattering}




\newcommand{\Rutgers}{Rutgers, The State University of New Jersey, Piscataway, New Jersey 08854, USA}
\newcommand{\Hampton}{Hampton University, Dept. of Physics, Hampton, VA 23668, USA}
\newcommand{\Dortmund}{Institute of Physics, Dortmund University, 44221, Germany }
\newcommand{\Otterbein}{Department of Physics, Otterbein University, 1 South Grove Street, Westerville, OH, 43081 USA}
\newcommand{\JMU}{James Madison University, Harrisonburg, Virginia 22807, USA}
\newcommand{\Florida}{University of Florida, Department of Physics, Gainesville, FL 32611}
\newcommand{\UCIrvine}{Department of Physics and Astronomy, University of California, Irvine, Irvine, California 92697-4575, USA}
\newcommand{\CBPF}{Centro Brasileiro de Pesquisas F\'{i}sicas, Rua Dr. Xavier Sigaud 150, Urca, Rio de Janeiro, Rio de Janeiro, 22290-180, Brazil}
\newcommand{\PUCP}{Secci\'{o}n F\'{i}sica, Departamento de Ciencias, Pontificia Universidad Cat\'{o}lica del Per\'{u}, Apartado 1761, Lima, Per\'{u}}
\newcommand{\INRM}{Institute for Nuclear Research of the Russian Academy of Sciences, 117312 Moscow, Russia}
\newcommand{\Jlab}{Jefferson Lab, 12000 Jefferson Avenue, Newport News, VA 23606, USA}
\newcommand{\Pittsburgh}{Department of Physics and Astronomy, University of Pittsburgh, Pittsburgh, Pennsylvania 15260, USA}
\newcommand{\Guanajuato}{Campus Le\'{o}n y Campus Guanajuato, Universidad de Guanajuato, Lascurain de Retana No. 5, Colonia Centro, Guanajuato 36000, Guanajuato M\'{e}xico.}
\newcommand{\Athens}{Department of Physics, University of Athens, GR-15771 Athens, Greece}
\newcommand{\Tufts}{Physics Department, Tufts University, Medford, Massachusetts 02155, USA}
\newcommand{\WM}{Department of Physics, College of William \& Mary, Williamsburg, Virginia 23187, USA}
\newcommand{\FNAL}{Fermi National Accelerator Laboratory, Batavia, Illinois 60510, USA}
\newcommand{\Purdue}{Department of Chemistry and Physics, Purdue University Calumet, Hammond, Indiana 46323, USA}
\newcommand{\MCLA}{Massachusetts College of Liberal Arts, 375 Church Street, North Adams, MA 01247}
\newcommand{\UMD}{Department of Physics, University of Minnesota -- Duluth, Duluth, Minnesota 55812, USA}
\newcommand{\Northwestern}{Northwestern University, Evanston, Illinois 60208}
\newcommand{\UNI}{Universidad Nacional de Ingenier\'{i}a, Apartado 31139, Lima, Per\'{u}}
\newcommand{\Rochester}{University of Rochester, Rochester, New York 14627 USA}
\newcommand{\Austin}{Department of Physics, University of Texas, 1 University Station, Austin, Texas 78712, USA}
\newcommand{\USM}{Departamento de F\'{i}sica, Universidad T\'{e}cnica Federico Santa Mar\'{i}a, Avenida Espa\~{n}a 1680 Casilla 110-V, Valpara\'{i}so, Chile}
\newcommand{\Geneva}{University of Geneva, 1211 Geneva 4, Switzerland}
\newcommand{\Chicago}{Enrico Fermi Institute, University of Chicago, Chicago, IL 60637 USA}
\newcommand{\hired}{}
\newcommand{\OregonState}{Department of Physics, Oregon State University, Corvallis, Oregon 97331, USA}
\newcommand{\bmeThanks}{now at SLAC National Accelerator Laboratory, Stanford, California 94309 USA}
\newcommand{\higueraThanks}{University of Houston, Houston, Texas, 77204, USA}
\newcommand{\damartinezThanks}{Now at Illinois Institute of Technology}
\newcommand{\joelmousseauThanks}{now at University of Michigan, Ann Arbor, MI, 48109}
\newcommand{\LazaThanks}{also at Department of Physics, University of Antananarivo, Madagascar}
\newcommand{\twaltonThanks}{now at Fermi National Accelerator Laboratory, Batavia, IL USA 60510}
\newcommand{\jwolcottThanks}{Now at Tufts University, Medford, Massachusetts 02155, USA }


\author{J.~Park}                          \affiliation{\Rochester}

\author{L.~Aliaga}                        \affiliation{\WM} \affiliation{\PUCP}
\author{O.~Altinok}                       \affiliation{\Tufts}
\author{L.~Bellantoni}                    \affiliation{\FNAL}
\author{A.~Bercellie}                     \affiliation{\Rochester}
\author{M.~Betancourt}                    \affiliation{\FNAL}
\author{A.~Bodek}                         \affiliation{\Rochester}
\author{A.~Bravar}                        \affiliation{\Geneva}
\author{H.~Budd}                          \affiliation{\Rochester}
\author{T.~Cai}                           \affiliation{\Rochester}
\author{M.F.~Carneiro}                    \affiliation{\CBPF}
\author{M.E.~Christy}                     \affiliation{\Hampton}
\author{J.~Chvojka}                       \affiliation{\Rochester}
\author{H.~da~Motta}                      \affiliation{\CBPF}
\author{S.A.~Dytman}                      \affiliation{\Pittsburgh}
\author{G.A.~D\'{i}az~}                   \affiliation{\Rochester}  \affiliation{\PUCP}
\author{B.~Eberly}\thanks{\bmeThanks}     \affiliation{\Pittsburgh}
\author{J.~Felix}                         \affiliation{\Guanajuato}
\author{L.~Fields}                        \affiliation{\FNAL}  \affiliation{\Northwestern}
\author{R.~Fine}                          \affiliation{\Rochester}
\author{A.M.~Gago}                        \affiliation{\PUCP}
\author{R.Galindo}                        \affiliation{\USM}
\author{A.~Ghosh}                         \affiliation{\CBPF} 
\author{T.~Golan}                         \affiliation{\Rochester}  \affiliation{\FNAL}
\author{R.~Gran}                          \affiliation{\UMD}
\author{D.A.~Harris}                      \affiliation{\FNAL}
\author{A.~Higuera}\thanks{\higueraThanks}  \affiliation{\Rochester}  \affiliation{\Guanajuato}
\author{J.~Kleykamp}                      \affiliation{\Rochester}
\author{M.~Kordosky}                      \affiliation{\WM}
\author{T.~Le}                            \affiliation{\Tufts}  \affiliation{\Rutgers}
\author{E.~Maher}                         \affiliation{\MCLA}
\author{S.~Manly}                         \affiliation{\Rochester}
\author{W.A.~Mann}                        \affiliation{\Tufts}
\author{C.M.~Marshall}                    \affiliation{\Rochester}
\author{D.A.~Martinez~Caicedo}\thanks{\damartinezThanks}  \affiliation{\CBPF}
\author{K.S.~McFarland}                   \affiliation{\Rochester}  \affiliation{\FNAL}
\author{C.L.~McGivern}                    \affiliation{\Pittsburgh}
\author{A.M.~McGowan}                     \affiliation{\Rochester}
\author{B.~Messerly}                      \affiliation{\Pittsburgh}
\author{J.~Miller}                        \affiliation{\USM}
\author{A.~Mislivec}                      \affiliation{\Rochester}
\author{J.G.~Morf\'{i}n}                  \affiliation{\FNAL}
\author{J.~Mousseau}\thanks{\joelmousseauThanks}  \affiliation{\Florida}
\author{D.~Naples}                        \affiliation{\Pittsburgh}
\author{J.K.~Nelson}                      \affiliation{\WM}
\author{A.~Norrick}                       \affiliation{\WM}
\author{Nuruzzaman}                       \affiliation{\Rutgers}  \affiliation{\USM}
\author{J.~Osta}                          \affiliation{\FNAL}
\author{V.~Paolone}                       \affiliation{\Pittsburgh}
\author{C.E.~Patrick}                     \affiliation{\Northwestern}
\author{G.N.~Perdue}                      \affiliation{\FNAL}  \affiliation{\Rochester}
\author{L.~Rakotondravohitra}\thanks{\LazaThanks}  \affiliation{\FNAL}
\author{M.A.~Ramirez}                     \affiliation{\Guanajuato}
\author{H.~Ray}                           \affiliation{\Florida}
\author{L.~Ren}                           \affiliation{\Pittsburgh}
\author{D.~Rimal}                         \affiliation{\Florida}
\author{P.A.~Rodrigues}                   \affiliation{\Rochester}
\author{D.~Ruterbories}                   \affiliation{\Rochester}
\author{H.~Schellman}                     \affiliation{\OregonState}  \affiliation{\Northwestern}
\author{C.J.~Solano~Salinas}              \affiliation{\UNI}
\author{N.~Tagg}                          \affiliation{\Otterbein}
\author{B.G.~Tice}                        \affiliation{\Rutgers}
\author{E.~Valencia}                      \affiliation{\Guanajuato}
\author{T.~Walton}\thanks{\twaltonThanks}  \affiliation{\Hampton}
\author{J.~Wolcott}\thanks{\jwolcottThanks}  \affiliation{\Rochester}
\author{M.Wospakrik}                      \affiliation{\Florida}
\author{G.~Zavala}                        \affiliation{\Guanajuato}
\author{D.~Zhang}                         \affiliation{\WM}

\collaboration{The MINER$\nu$A Collaboration}\ \noaffiliation

\date{\today}

\pacs{13.15.+g,13.66-a}
\begin{abstract}
Muon-neutrino elastic scattering on electrons is an observable neutrino process whose
cross section is precisely known. Consequently a measurement of this
process in an accelerator-based $\nu_\mu$ beam can improve the knowledge of the absolute neutrino 
flux impinging upon the detector; typically this knowledge is limited
to $\sim10\%$ due to uncertainties in hadron production and focusing.
We have isolated a sample of $135\pm17$ neutrino-electron elastic scattering candidates in the segmented scintillator
detector of MINERvA, after subtracting backgrounds and correcting for efficiency. We show how this sample
can be used to reduce the total uncertainty on the NuMI $\nu_\mu$ flux from 9\% to 6\%. 
Our measurement provides a flux constraint that is useful to other experiments using the
NuMI beam, and this technique is applicable to future neutrino beams operating at multi-GeV energies.

\end{abstract}
\ifnum\sizecheck=0  
\maketitle
\fi

\section{Introduction}
\label{sec:intro}

Neutrino-electron
elastic scattering is precisely predicted in the electroweak Standard
Model because it involves only the scattering of fundamental leptons.
At tree level and in the limit that the neutrino energy $E_\nu$ 
is much greater than the electron mass
$m_e$, 
the $\nu e \rightarrow \nu e$ cross section for all active neutrinos 
and antineutrinos is given generically by
\begin{equation}
\frac{d\sigma(\nu e^-\to\nu e^-)}{dy} = \frac{G^2_F s}{\pi}\left[
  C_{LL}^2+C_{LR}^2(1-y)^2\right] ,
\label{eqn:tree-xsec}
\end{equation}
where $G_F$ is the Fermi weak coupling constant, $s$ is the Mandelstam
invariant representing the square of the total energy in the center-of-mass frame, and $y\equiv T_e/E_\nu$ where $T_e$ is the electron kinetic energy.  The couplings $C_{LL}$ and $C_{LR}$
depend on the neutrino flavor and whether the incident particle is a
neutrino or antineutrino.  For $\nu_\mu$ and $\nu_\tau$, 
$C_{LL}=\frac{1}{2}-\sin^2\theta_W$ and $C_{LR}=\sin^2\theta_W$, where
$\theta_W$ is the Weinberg angle, and for the corresponding antineutrinos the values
for $C_{LL}$ and $C_{LR}$ are interchanged.  For $\nu_e$ ($\bar\nu_e$), the value of $C_{LL}$ ($C_{LR}$) is 
$\frac{1}{2}+\sin^2\theta_W$ because the interaction contains
interfering contributions from the neutral-current interaction that is
present for all flavors and from a charged-current interaction that is present only
for electron neutrinos.  The kinematics of the reaction limit the
magnitude of the four-momentum transferred from the neutrino, $q$, to be less than $\sqrt{s}$.  
The angle of the final state electron with respect to the neutrino, $\theta$, is
uniquely determined from the initial neutrino and final lepton energies
by
\begin{equation}
1-cos\theta = \frac{m_e(1-y)}{E_e};
\label{eqn:angle}
\end{equation} 
therefore at accelerator neutrino energies, where $m_e\ll E_\nu$, the
final state electron is very forward.
Electroweak radiative corrections for these
cross sections have been calculated to one loop~\cite{Sarantakos:1982bp} and constitute few-percent corrections to the
tree-level expressions for GeV-energy neutrinos.  The prediction can be 
further improved by including additional low-energy
terms due to radiative corrections~\cite{Bahcall:1995mm} and one-loop electroweak
couplings from recent global fits to electroweak data~\cite{Erler:2013xha}.

Experimental measurements of $\nu_\mu e^-$ and $\nubar_\mu e^-$
elastic scattering have been performed by the CHARM experiment at CERN~\cite{nue-CHARM}, 
the E734 experiment at Brookhaven~\cite{nue-734}  and, most precisely, by the
CHARM-II experiment at CERN~\cite{nue-CHARM2}.  In addition, $\nu_e e^-$ scattering has
been studied by the E-225 and LSND
experiments at LAMPF~\cite{nue-LAMPF-E-225,nue-LSND}, and $\nubar_e e^-$ scattering by the
TEXONO experiment~\cite{nue-TEXONO}.  These measurements are limited
in precision either by statistics of the neutrino-electron elastic
scattering sample, or by knowledge of the incoming neutrino flux, or both.

The theoretical uncertainty of the neutrino-electron scattering cross section is much smaller than the uncertainty associated with any one or the combination of all measurements~\cite{Erler:2013xha}.
This unusual situation in neutrino scattering
allows the use of this process as a standard candle from which one 
can derive constraints on the neutrino flux.  
Given the above equations, the $\nu_e$ ($\nu_\mu$ or $\nu_\tau$) cross section varies by only 15\% (20\%) as a function of $y$, however, and therefore the energy of the final state electron is only loosely correlated with the energy of the incoming neutrino.  The total number of electron scattering events provides a strong constraint on the integral of the flux, and the electron energy distribution itself provides only a small additional constraint on the neutrino energy spectrum.

The technical challenge
that offsets this advantage is that from Eqn.(\ref{eqn:tree-xsec}) the cross section is small, roughly
$10^{-4}$ of the total charged-current $\nu_\mu$ cross section, 
meaning signal statistics
are low and backgrounds substantial.  However, with an intense neutrino beam and a capable detector, the statistical precision of the neutrino-electron scattering measurement may rival or exceed that of the flux prediction.  An {\em in situ} measurement has the added benefit that it accounts for all effects of the beam optics such as horn current and geometry which can be difficult to predict precisely.  

The Neutrinos at the Main Injector (NuMI) beamline at Fermilab~\cite{Adamson:2015dkw} has a flux prediction whose precision is limited primarily by uncertainties in the energy and angular spectra of hadrons produced by the incoming proton beam.  For the configuration represented by the measurement reported here, the NuMI flux prediction uncertainty ranges from 10\% at the 3~GeV peak of the flux, to significantly higher uncertainties above the peak~\cite{nubarprl,Adamson:2007gu}.  At the same time, the high intensity of the NuMI beamline means that the total signal sample in a neutrino-electron elastic scattering analysis in a multi-ton detector has comparable statistical precision to the flux uncertainty.  

This article describes a measurement of neutrino-electron scattering using the 6-ton MINERvA scintillator tracking detector.  Specifically, the number of these events in the MINERvA detector is measured as a function of electron energy and used to constrain the uncertainty on the NuMI beam flux, which consists primarily of $\nu_\mu$. 
The signature for neutrino-electron scattering
is a single electron with energy and angle satisfying  $E_e\theta^2<2m_e$, given Eqn.~\ref{eqn:angle},  and no other activity in the event.  
The dominant
backgrounds come from electrons produced in charged
current $\nu_e$ and $\nubar_e$ interactions, and decay photons from $\pi^0$ production.  Therefore, the
analysis selects low angle electrons, rejects photons, and rejects 
events with any
other particles visible in the detector.

\section{\minerva Experiment and Data}
\label{sec:expt}

\begin{figure}[tp]
\centering
\ifnum\PRLsupp=0
  \includegraphics[width=\columnwidth]{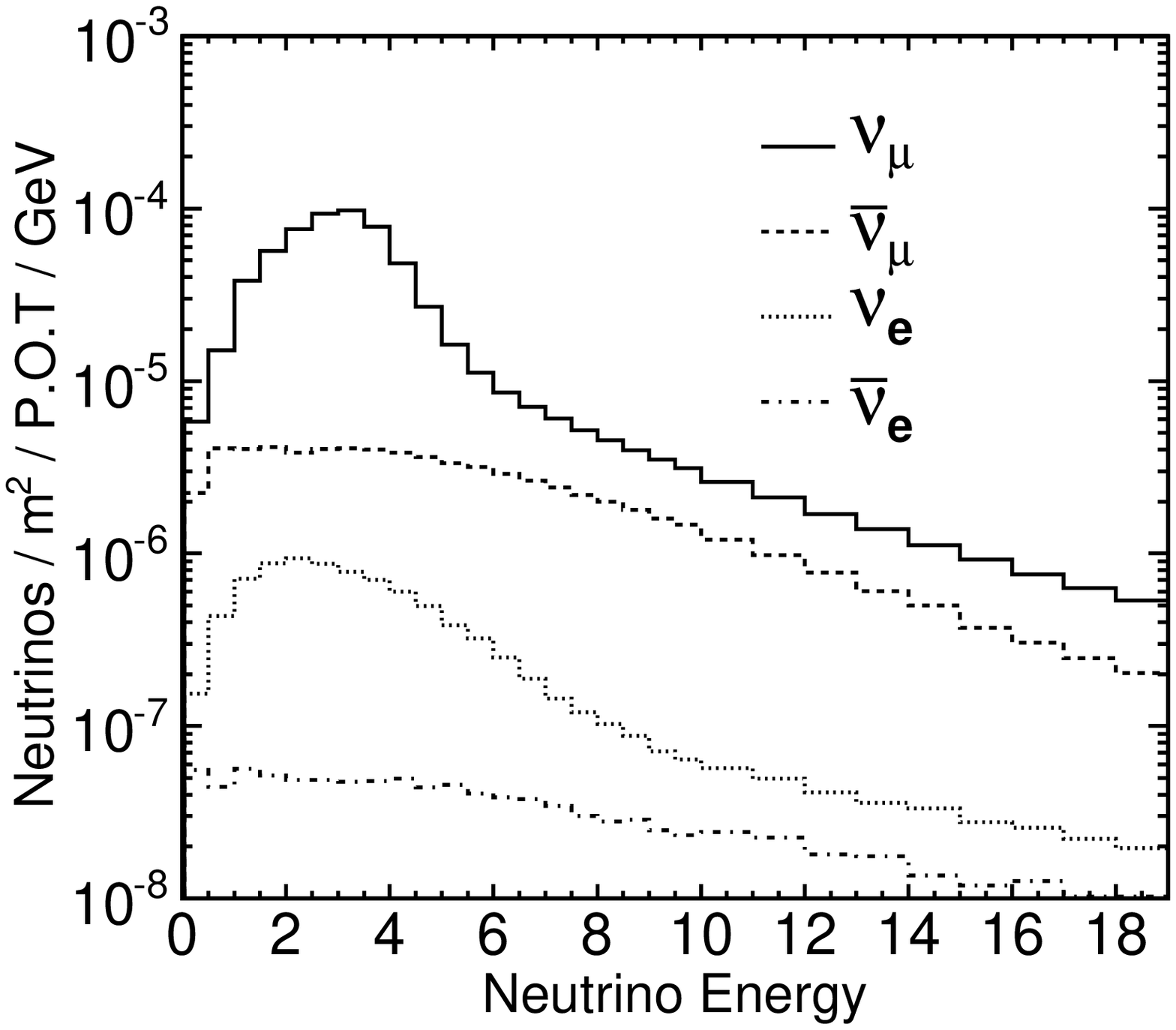 }
\else
  \includegraphics[width=0.8\columnwidth]{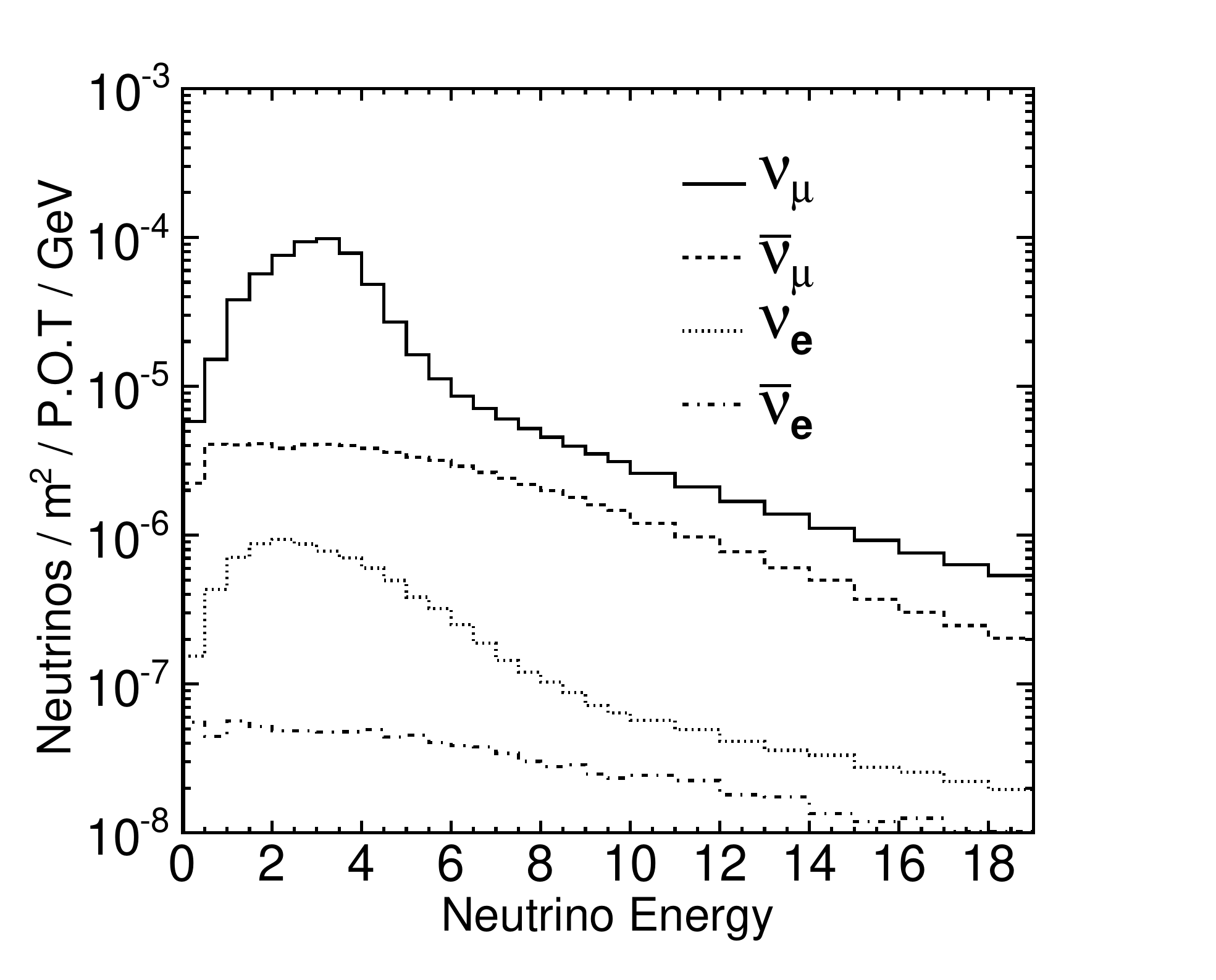} 
\fi
\caption{The predicted flux of $\nu_\mu$, $\nubar_\mu$, $\nu_e$ and $\nubar_e$ 
for the data used in this analysis.}
\label{fig:flux}
\end{figure}

The \minerva experiment uses the NuMI
beam~\cite{Adamson:2015dkw}, which begins with \unit[120]{GeV} protons 
striking a graphite target.  The mesons produced in $p+C$ interactions
are focused by two magnetic horns into a \unit[675]{m} 
helium-filled decay pipe.  For the data presented here, the horns are set to focus positive
mesons, resulting in a $\nu_\mu$-enriched beam whose peak energy is 3~GeV.  Muons produced in
meson decays are absorbed in \unit[240]{m} of rock downstream of the
decay pipe. This analysis uses data taken between November 2010 and
April 2012 with $3.43\times 10^{20}$ protons on target.  The predicted
flux of neutrinos for this exposure is shown in Fig.~\ref{fig:flux}, and integrated over
all energies, the beam is 
92.9\% $\nu_\mu$, 5.8\% $\bar\nu_\mu$ and 1.3\% ($\nu_e+\bar\nu_e$).  

The \minerva detector consists of a core of scintillator
strips surrounded by electromagnetic and hadronic calorimeters 
(ECAL and HCAL, respectively) on the
sides and downstream end of the detector.  The \minerva
scintillator tracking region is composed of 95\% CH and 5\% other materials by
weight~\cite{minerva_nim}.  The strips are perpendicular to the
z-axis (which is horizontal and approximately aligned with the beam axis) and are arranged in planes
with a 1.7~cm strip-to-strip pitch\footnote{The y-axis points along
the zenith and the beam is directed downward by \unit[58]{mrad} in the
y-z plane.}. Three plane orientations ($0^\circ, \pm 60^\circ$
rotations around the z-axis, denoted $X, U,$ and $V$) facilitate 
three-dimensional reconstruction of the neutrino
interaction vertex and of outgoing charged particle tracks.  
 The \unit[3.0]{ns}
timing resolution of the detector allows separation of particles from multiple interactions within a
single beam pulse.  
\minerva is located \unit[2]{m} upstream of the MINOS near detector, 
a magnetized iron-scintillator tracking spectrometer~\cite{Michael:2008bc}.  Although the latter detector is not used
directly in this analysis, it is used to reconstruct the momentum of
through-going muons for many calibrations~\cite{minerva_nim} and to perform reconstruction efficiency studies, as described in paragraphs below.

\section{Experiment Simulation}
\label{sec:exptsim}

The neutrino beam is simulated by a Geant4-based
model~\cite{Agostinelli2003250,1610988} which is constrained by 
NA49 proton+carbon hadron production measurements~\cite{Alt:2006fr}.  
FLUKA is used to shift NA49
measurements to match proton energies in NuMI, which range from the primary proton energy of \unit[120]{GeV} down to secondary proton energies of  
\unit[12]{GeV} ~\cite{Ferrari:2005zk,Battistoni:2007zzb}.  The
$\pi / K$ ratio measured by MIPP on a thin carbon
target~\cite{Lebedev:2007zz} is used to constrain production of kaons.
Hadronic interactions not constrained by the NA49 or MIPP data are predicted
using the FTFP\_BERT hadron shower model implemented in Geant4 version 9.2 patch 3. 

Neutrino interactions are simulated using the GENIE 2.6.2 neutrino
event generator~\cite{Andreopoulos201087}.  GENIE provides the tree-level neutrino-electron scattering cross section described above, which we modify to account for next-to-leading-order 
radiative corrections as described \SuppLocation.  For quasielastic $\nu_e$
interactions, the cross section is given by the Llewellyn Smith
formalism~\cite{LlewellynSmith:1971zm}.  Vector form factors come from
fits to electron scattering data~\cite{Bradford:2006yz}; the axial
form factor used is a dipole with an axial mass ($M_A$) of
0.99~GeV$/$c$^2$, consistent with deuterium
measurements~\cite{Bodek:2007vi,Kuzmin:2007kr}, and sub-leading form
factors are assumed from PCAC or exact G-parity
symmetry~\cite{Day:2012gb}.  The nuclear model is the relativistic
Fermi gas with a Fermi momentum of
$221$~MeV$/$c and with an extension to higher nucleon momenta to account
for short-range
correlations~\cite{Bodek:1980ar,Bodek:1981wr}.  Inelastic
reactions with a low hadronic invariant mass are based on a tuned model of discrete baryon resonance
production~\cite{Rein:1980wg}, and the transition to deep inelastic
scattering is simulated using the Bodek-Yang
model~\cite{Bodek:2004pc}. Final state interactions
are modeled using the INTRANUKE
package~\cite{Andreopoulos201087}.  Coherent pion production is
simulated using the model of Rein and
Sehgal~\cite{Rein:1982pf}.  Uncertainties in the parameters
of these models are assigned according to uncertainties
in experimental measurements or to cover differences between experiments and
model predictions.

The \minerva detector's response is simulated by a tuned
Geant4-based~\cite{Agostinelli2003250,1610988} program, version 9.4 patch 2, with the QGSP\_BERT hadron cascade model.  The energy
scale of the detector is set by ensuring that both the photostatistics
and the reconstructed energy deposited by momentum-analyzed
through-going muons agree in data and simulation.  The calorimetric
constants used to reconstruct the energy of electromagnetic showers
and correct for passive material are determined from the
simulation.  The uncertainty in the detector's response to protons and charged
pions is constrained by the measurements made with a scaled-down
version of the \minerva detector in a low energy hadron test
beam~\cite{testbeam_nim}.  The energy scale for electrons in
the scintillator tracker is verified using a sample of Michel
electrons from $\mu^\pm\to e^\pm\nu\nubar$ decays of muons stopping in
the detector~\cite{minerva_nim}, by the reconstructed invariant
mass of identified $\pi^0\to\gamma\gamma$ decays~\cite{Trung-pi0}, and 
in test beam electron measurements~\cite{testbeam_nim}.


\section{Event Reconstruction}
\label{sec:reco}

The \minerva\ detector records the energy and time of energy depositions (hits) in each scintillator bar.  Hits are first grouped in time and then clusters of energy are formed by spatially grouping the hits in each scintillator plane.  Clusters with energy $>\unit[1]{MeV}$ are then matched among the three views to create a track. 
An electron typical of a $\nu e\rightarrow \nu e$ scatter features a single particle track near the neutrino interaction vertex at which it was created, gradually developing into an electromagnetic cascade in the scintillator and terminating in the downstream electromagnetic calorimeter.  
An energetic electron typically traverses at least one radiation length as a minimum-ionizing particle (MIP) until it begins to shower. The 40~cm radiation length of scintillator corresponds to 25 planes when the direction of the electron is normal to the planes.  The MIP-like segment can be identified as a track, and the beginning (angle) of that track serves as the event vertex (electron angle).  Occasionally, an electron starts to shower early and the MIP track is too short to be reconstructed as a track.  In this case the topologically contiguous energy deposition is used in a least-squares fit to define the vertex location and shower direction which are inputs to the cone algorithm described below.  
%
%
Only events with an event vertex within the central 112 planes of the scintillator tracking region and no closer than \unit[4]{cm} to any edge of the central tracker are retained as signal candidates. These requirements define a region with a mass of \unit[6.10]{metric tons}.

Once a track or an isolated energy deposition is identified, a search cone is formed using the vertex and angle of the identified object.  The cone is defined to have an opening angle of 10 degrees with respect to the electron direction, and it begins at a location upstream of the vertex such that the width of the cone 80~mm upstream of the vertex is 50~mm.  
The cone extends far enough to capture the downstream remnants of the electromagnetic showers which sometimes fluctuate to only a single photon which later converts.  
The energy within the search cone is summed according to the calorimetric tuning as defined above; this is identified as the electron candidate energy.  The resulting electron fractional energy resolution using this procedure is   $5.9\%/\sqrt{E_e /GeV} \oplus 3.4\%$~\cite{Park:2013dax}.  

Accurate reconstruction of the electron shower direction is critical to the rejection of backgrounds using an $E_e\theta^2$ cut, as discussed below. The energies and locations of clusters inside the cone are fed into a Kalman filter to determine the electron angle with respect to the beam direction.  Because the downstream end of an electron shower does not necessarily align with the original electron direction, only the most upstream 30 clusters are used in the fit.  The resulting average electron angular resolution is 7.2 (7.5)~mrad in the horizontal (vertical) direction~\cite{Park:2013dax}.   

The event interaction time is inferred from the times of the tracked hits.  
Other (untracked) clusters that are within \unit[20]{ns} before and 
\unit[35]{ns} after that time are also associated with the event.  Energy within this reconstruction time window, but outside the electron cone, is used to search for the presence of other particles in the event which would indicate that the event is a background rather than neutrino-electron elastic scattering event.  

\section{Event Selection}
\label{sec:selection}

The majority of neutrino interactions in \minerva\ come from charged-current (CC) $\nu_\mu$ interactions on nuclei either in or upstream of the detector.  Events truly originating upstream of the fiducial volume but reconstructed within it can be rejected simply by requiring the energy in a 30 cm-diameter cylinder along the cone axis and upstream of the reconstructed event vertex be less than 300~MeV.  True fiducial $\nu_{\mu}$ CC events, on the other hand, can be identified by the presence of a muon or other MIP-like charged particle, which will frequently penetrate through the ECAL into the HCAL, in contrast to electron showers, which typically end in the ECAL. 
Events are removed if the end of the shower penetrates through more than 2~planes of the hadron calorimeter, which corresponds to 5~cm of steel and 3~cm of scintillator, or 3 radiation lengths for normally incident particles.  

After the $\nu_\mu$ CC interactions on nuclei are removed, the remaining background is from neutral-current (NC) pion production or electron neutrino interactions on nuclei in the detector.  These topologies are removed with a series of cuts described below.  

A minimum energy of 0.8~GeV is required to remove the significant background that arises from $\pi^0$ decays to photons and to ensure good angular and energy reconstruction of the electron.  Given the dependence of the cross section on the electron energy, this cut is 45\% (50\% ) efficient for $\nu_\mu$'s ($\nu_e$'s) at 2~GeV, and rises by $0.8GeV/E_\nu$ to 85\% (90\% ) efficient at 10~GeV.  

In addition, the electron track is not allowed to bend by more than 9 degrees, since this would indicate a hadronic scatter.  
To ensure that the search cone contains the energy of only one particle, cuts are made on the transverse and longitudinal energy distributions and on the consistency of the energy depositions between the three views of the scintillator planes.

Two transverse energy cuts are made to remove two-particle backgrounds.  These cuts were set using the simulation and optimizing the cut such that the most background was removed while still retaining signal efficiency.  
Firstly, for electrons that are less than 7~GeV in energy, the energy within 5~cm of the outer boundary of the cone is required to be less than 120~MeV.  For those electrons with energies above 7~GeV, that cut is relaxed and the energy in that same region is required to be less than $120+7.8\times (E_e/GeV-7)$~MeV.  

\begin{figure}[tp]
\centering
\ifnum\PRLsupp=0
\includegraphics[width=\columnwidth]{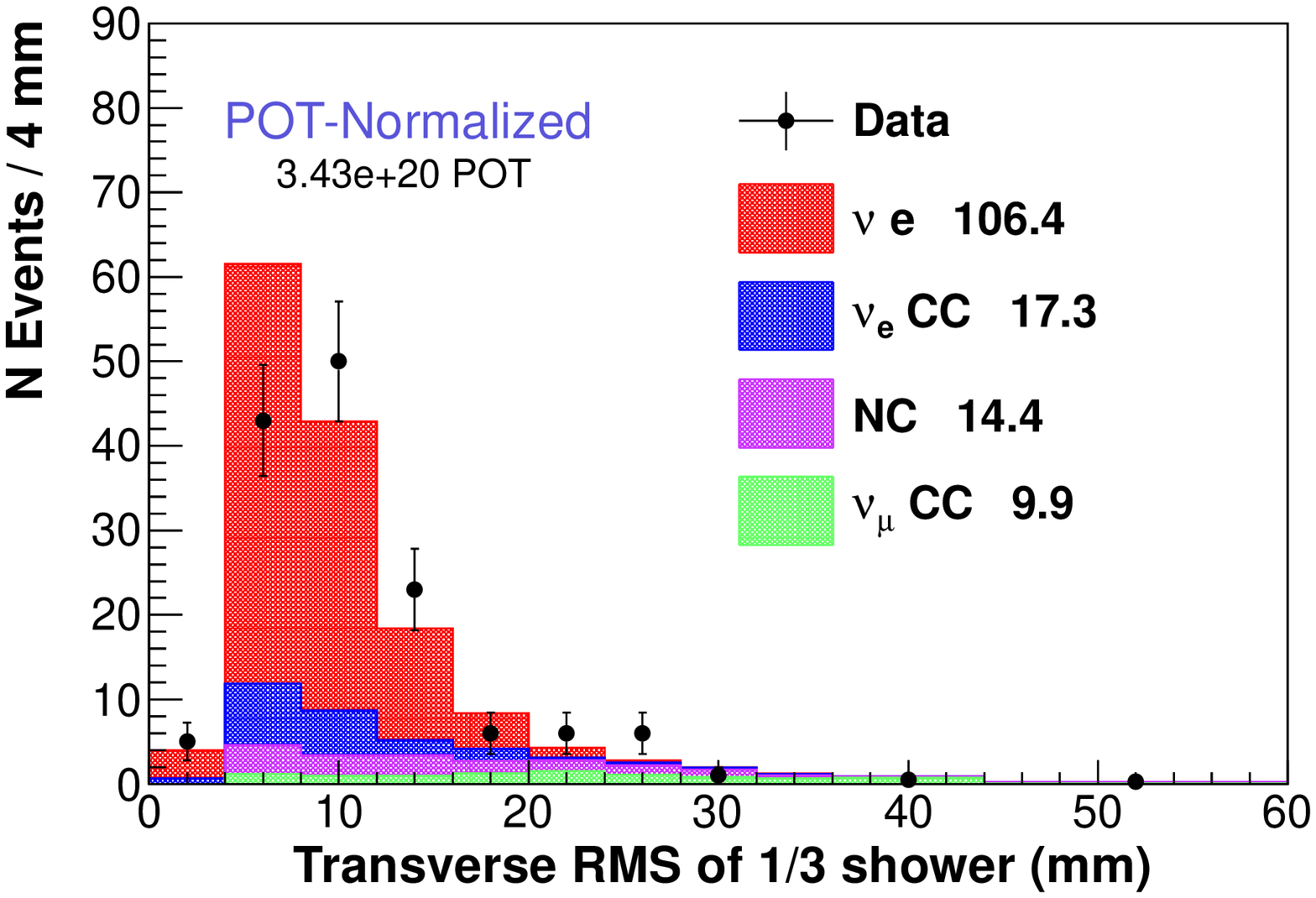}
\else
  \includegraphics[width=0.8\columnwidth]{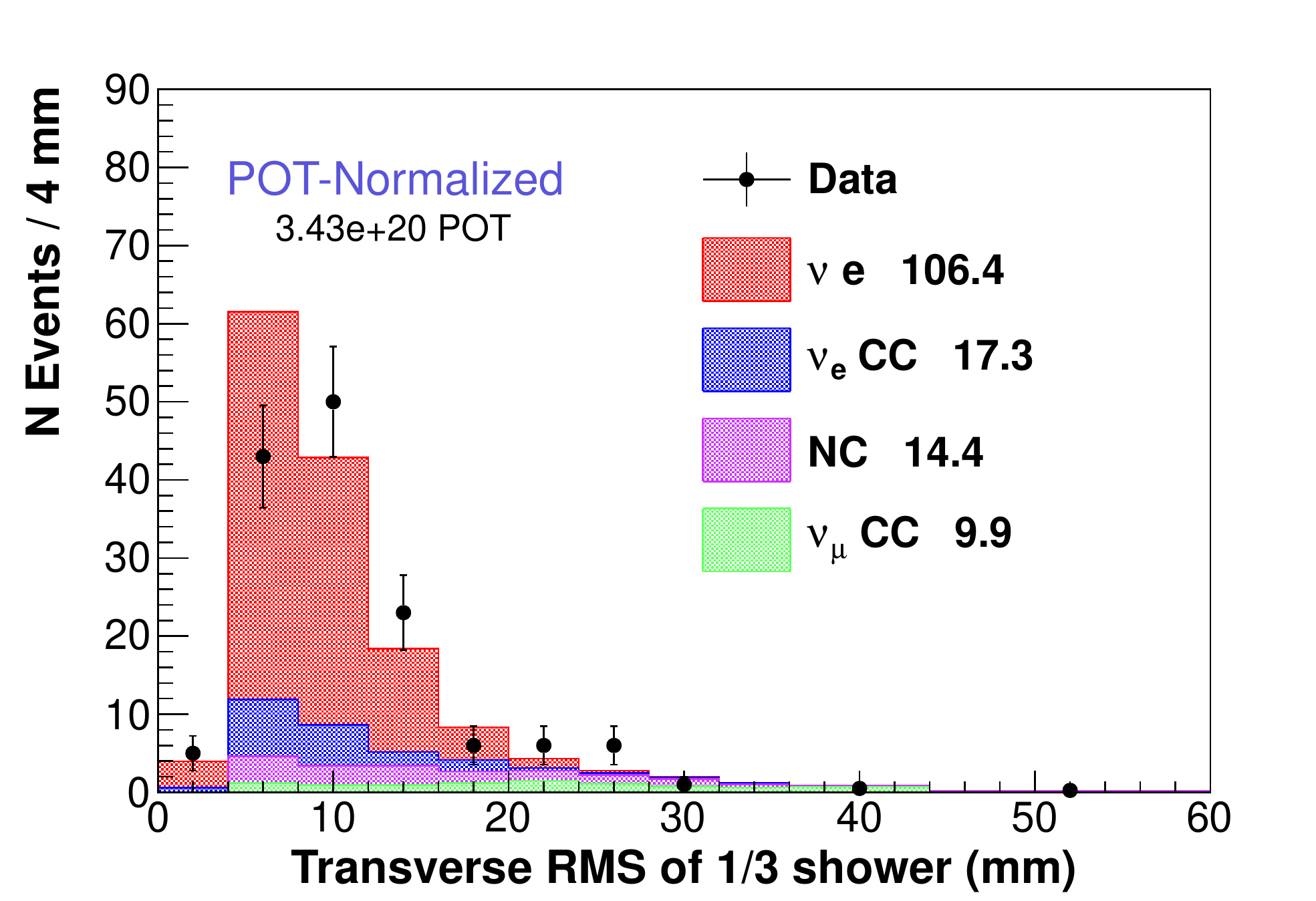}
\fi
\caption{The energy-weighted RMS of the transverse distance of each cluster of energy from the cone center, averaged over the planes in the first third of the shower for data and simulation, after the simulated backgrounds have been tuned.   The simulation has been divided according to channel into $\nu_\mu$ and $\nu_e$ scattering on electrons ($\nu$ e), $\nu_e$ charged current interactions ($\nu_e$ CC), other neutral current interactions (NC), and $\nu_\mu$ charged current interactions ($\nu_\mu$ CC).  The numbers in the legend denote the total number of simulated events in each channel after background tuning.  }
\label{fig:trans_rms}
\end{figure}

Secondly, for each view, the energy-weighted RMS of the distances of each cluster from the cone center in the first third of the shower must be less than \unit[20]{mm}.  The distribution of events for data and simulation, after the background tuning discussed below and after all cuts but this one are made, is shown in Fig.~\ref{fig:trans_rms}.  

Cuts are also made on the longitudinal energy distribution to ensure that the shower is from a single electromagnetic particle.     
The Kalman filter that determines the electron angle returns a $\chi^2$ describing the quality of the fit to a single-particle energy deposition.  A very loose cut requiring the $\chi^2$ per number of degrees of freedom to be less than 100 is made to remove multiple particle showers without compromising the single-particle acceptance.  In addition, the longitudinal position of the plane containing the maximum energy deposition must be at a distance from the shower start that is consistent with electromagnetic shower propagation in scintillator.  This removes two-photon events since the two photons, even if overlapping in space, will not usually convert at the same point.  

Finally, the energy deposition in the search cone for each view relative to the other two views is required to be consistent with that of a single particle.  When there are two or more particles originating from  the same vertex, they will rarely overlap in more than one detector view.  Because there are twice as many planes in the $X$ orientation as in the $U$ or $V$ orientation, the following two cuts remove events where two or more particles overlap inside the cone in one view but not all views:  
\begin{eqnarray*}
 \left| \frac{E_X-E_U-E_V}{E_X + E_U + E_V}  \right|  &<&  0.28 ,  \\
\left| \frac{E_U-E_V}{E_U + E_V}  \right|  &<&  0.5 .
\end{eqnarray*}
Here, $E_J$ is the energy deposited in the $J$ plane orientation of the detector.  

\begin{figure}[tp]
\centering
\ifnum\PRLsupp=0
  \includegraphics[width=\columnwidth]{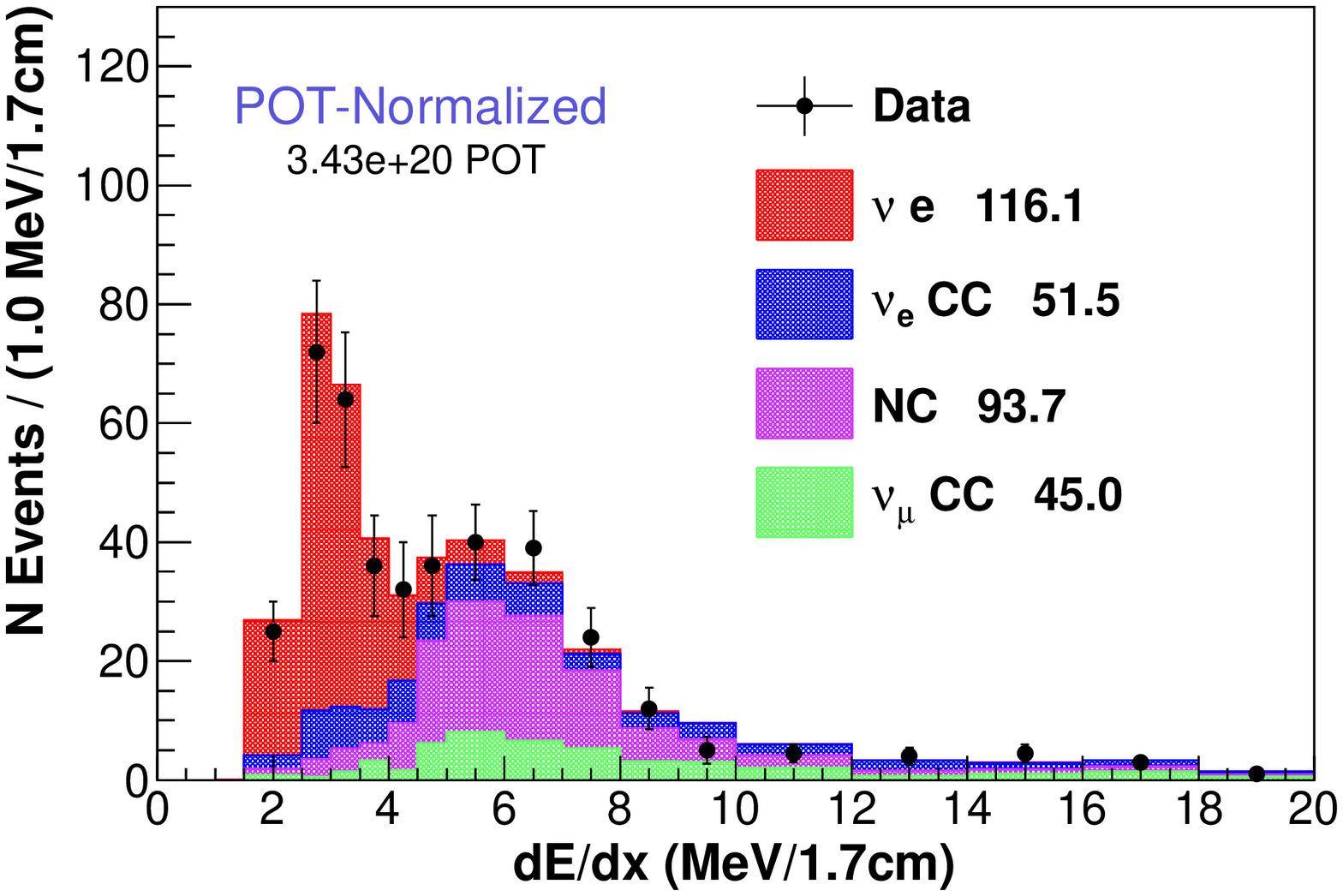}
\else
  \includegraphics[width=0.8\columnwidth]{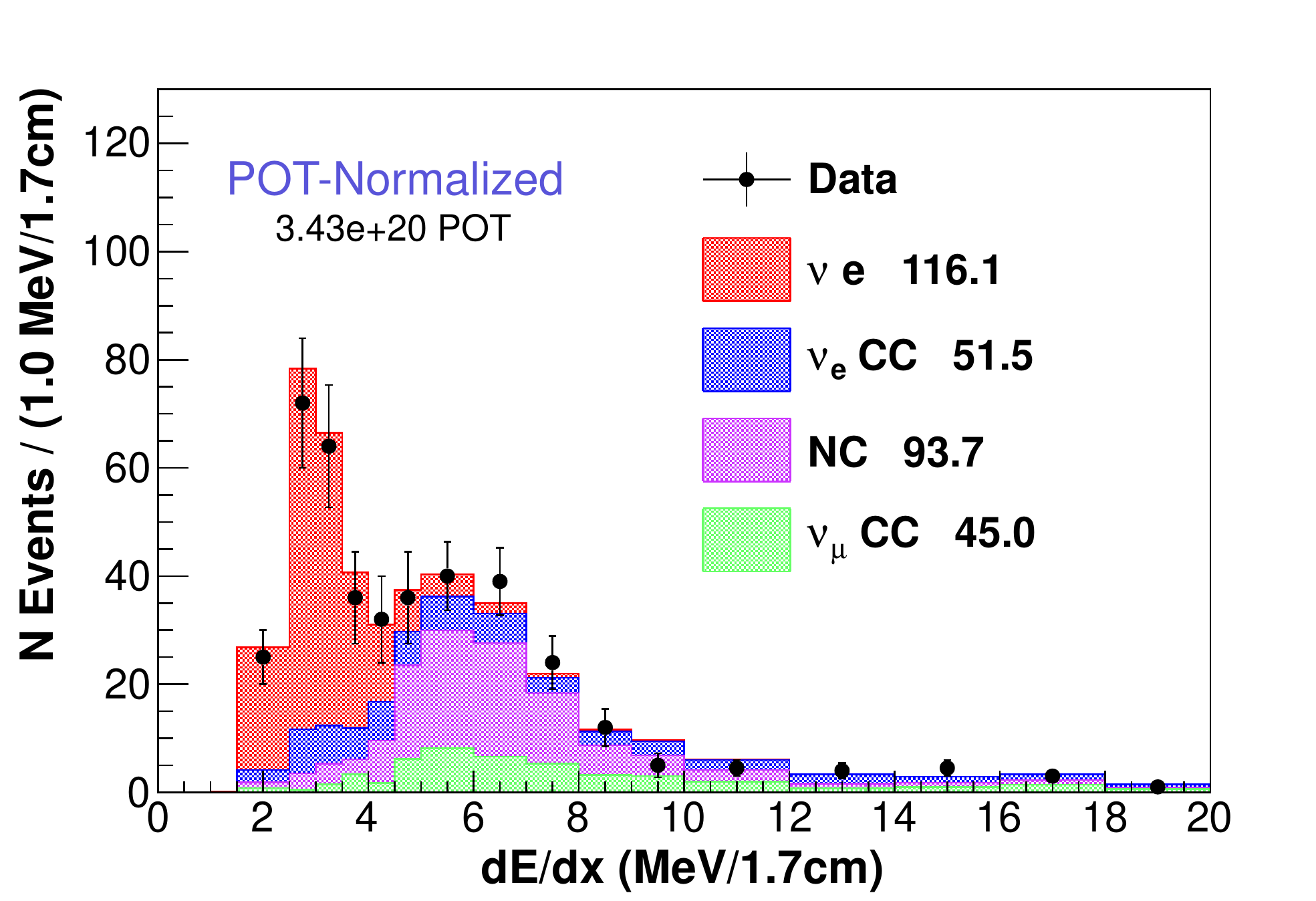} 
\fi
\caption{The distribution of $dE/dx$ for data and simulation after the cuts that isolate single electromagnetic showers are made and after the backgrounds are tuned, but before the final $dE/dx$ cut is made.  
The simulation has been divided according to channel into $\nu_\mu$ and $\nu_e$ scattering on electrons ($\nu$ e), $\nu_e$ charged current interactions ($\nu_e$ CC), other neutral current interactions (NC), and $\nu_\mu$ charged current interactions ($\nu_\mu$ CC).  The numbers in the legend denote the total number of simulated events in each channel after background tuning.  
Signal candidates are required to have an average $dE/dx$ less than 4.5 MeV $/$ 1.7 cm.}  
\label{fig:dedx}
\end{figure}

After the cuts above, there are still 32 thousand events with fewer than 200 signal events expected.  The remaining backgrounds are primarily from $\nu_e$ quasielastic interactions, and single-photon events. 
Photons can be rejected by looking at the energy deposition per unit distance ($dE/dx$) at the beginning of the electron candidate track.  For photons that convert, $dE/dx$ is consistent with that made by two electrons while the signal $dE/dx$ is that of only one electron.  The cut is best made before the electron starts showering, but far enough into the track that the photostatistics are adequate.  
The optimal distance for this analysis is to cut on the average energy deposition in the first four scintillator planes of the track.  This average energy deposition, normalized by the cosine of the incident electron, is shown for data and predicted signal and background events in Fig.~\ref{fig:dedx}.  Signal events are required to have an average $dE/dx$ less than 4.5 MeV $/$ 1.7 cm, where 1.7~cm corresponds to the thickness of one scintillator plane.  

\begin{figure}[tp]
\centering
\ifnum\PRLsupp=0
  \includegraphics[width=\columnwidth]{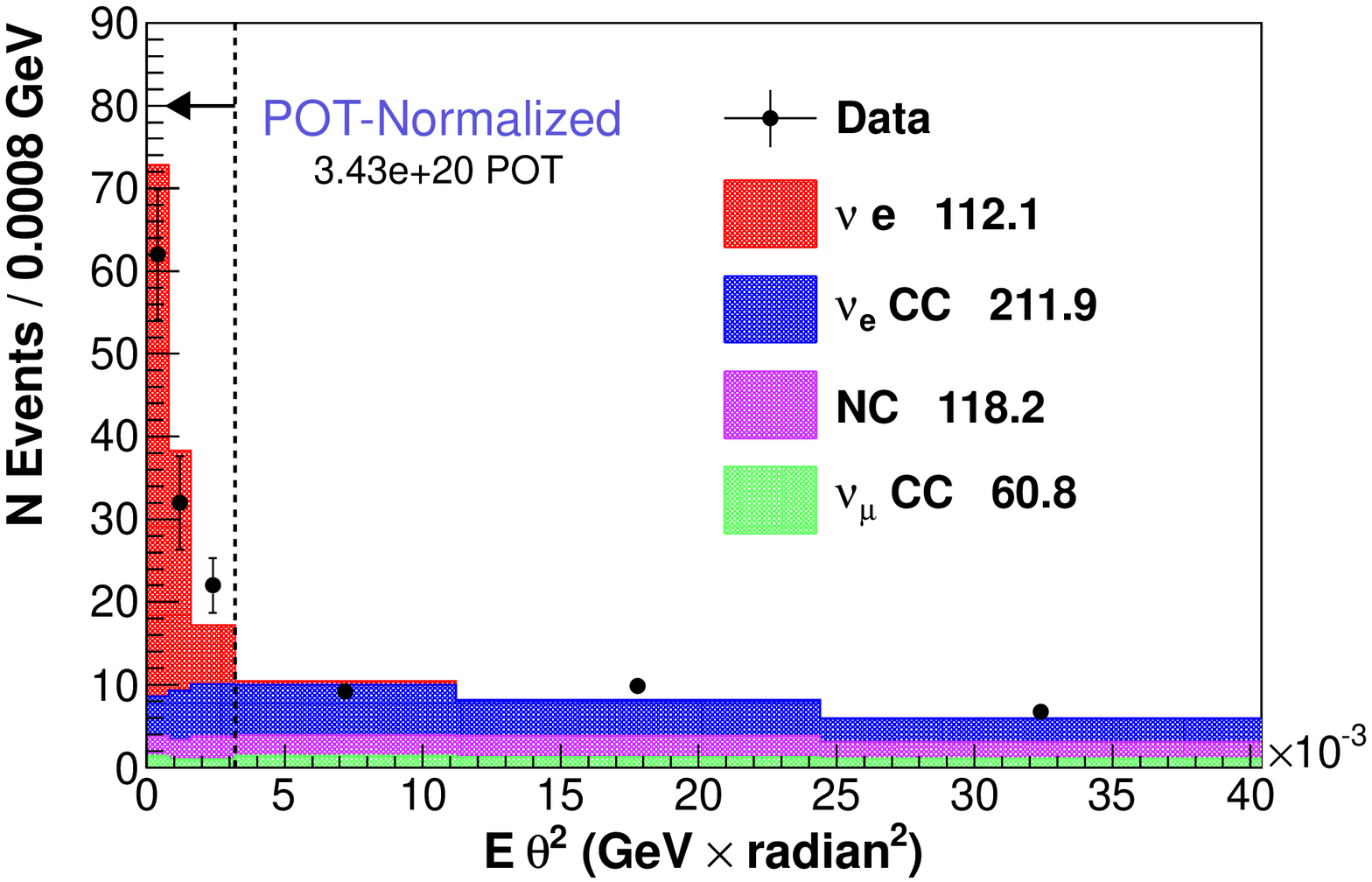}
\else
  \includegraphics[width=0.8\columnwidth]{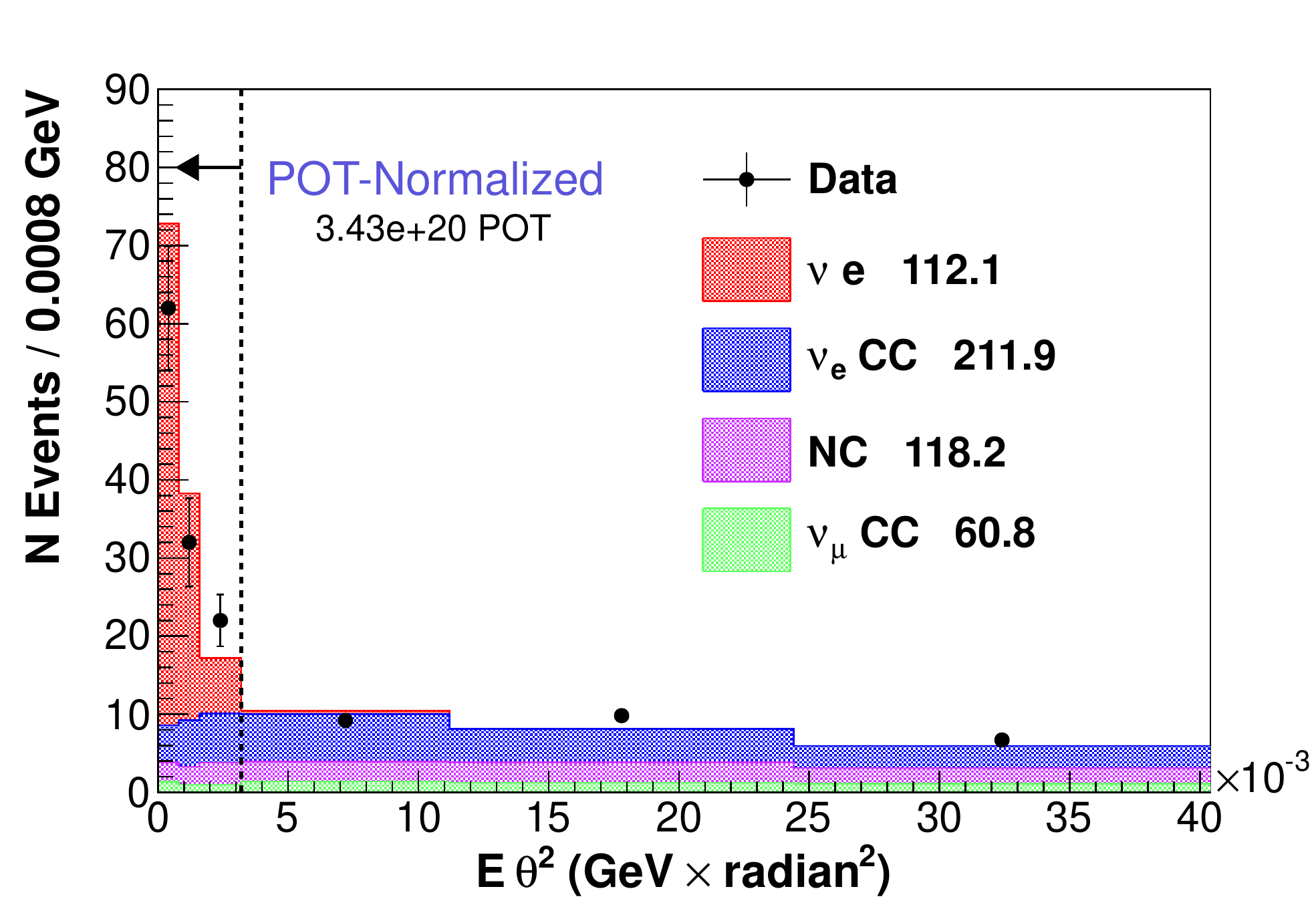} 
\fi
\caption{The distribution of $E_e \theta^2$ for data and simulation after the backgrounds are tuned and after all cuts except the $E_e\theta^2$ cut are made.  The simulation has been divided according to channel into $\nu_\mu$ and $\nu_e$ scattering on electrons ($\nu$ e), $\nu_e$ charged current interactions ($\nu_e$ CC), other neutral current interactions (NC), and $\nu_\mu$ charged current interactions ($\nu_\mu$ CC).  The numbers in the legend denote the total number of simulated events in each channel after background tuning.  The signal region is defined as events with $E_e\theta^2$ less than 0.0032~GeV$\times$radian$^2$.  }
\label{fig:eth2}
\end{figure}

After the $dE/dx$ cut is made the remaining major background is from  $\nu_e$ charged current quasi-elastic interactions (CCQE), namely $\nu_e n \rightarrow e^- p$ or $\bar\nu_e p \rightarrow e^+ n$.  If the recoiling nucleon is not observed in the detector as is common at low momentum transfer, the final
state is a single electron or positron, which cannot be distinguished from the signal using particle identification cuts.  Given the kinematics described by Eqn.~\ref{eqn:angle} and the small angle approximation, $E_e\theta^2$ must be less than the electron mass for neutrino-electron scattering, but is usually much larger for neutrino-nucleon scattering.  
Figure~\ref{fig:eth2} shows the distribution of this quantity for the data, and the signal and background predictions, after all cuts except the $E_e\theta^2$ cut.  Events with $E_e\theta^2$ greater than \unit[0.0032]{GeV radian}$^2$ are removed.  

The $E_e\theta^2$ cut removes the $\nu_e$ CCQE background effectively at low energy, but it is less effective for high energy electrons because those electrons are also produced at smaller angles, as in neutrino-electron scattering.  An additional cut on the momentum transfer squared, $Q^2$, reconstructed directly under the assumption of $\nu_e$ CCQE kinematics, is applied, where 
\begin{eqnarray}
E_\nu &=& \frac{m_nE_e - m_e^2/2}{m_n - E_e + P_e\cos\theta} , \\
Q^2 &=& 2m_n(E_\nu - E_e),   
\end{eqnarray}
where $m_n$ is the neutron mass.  Events with $Q^2$ less than \unit[0.02]{GeV}$^2$ are removed to reject high energy electron $\nu_e$ CCQE events.  This cut is 98\% efficient for signal and removes 30\% of the electron neutrino CCQE background.

\section{Background Subtraction}
\label{sec:background}

As shown in Fig.~\ref{fig:eth2}, the number of predicted background events after the final event selection is a small fraction of the signal events. 
To produce the signal electron energy distributions, the backgrounds must be estimated and subtracted.  
This procedure is subject to systematic uncertainties because mis-modeling of both the background and the neutrino flux can bias the signal measurement.   

To reduce the background prediction uncertainty and the dependence of the backgrounds on the {\em a priori} flux prediction, the analysis normalizes the background prediction using events that fail the $E_e\theta^2$ cut but still pass a loose $dE/dx$ cut.  
The sideband is defined to be all events with $E_e\theta^2$ greater than \unit[0.005]{GeV radian}$^2$ and $dE/dx$ less than \unit[20]{MeV/1.7cm}.  This region is chosen with a sufficiently high $E_e\theta^2$ value so that it contains no signal events but does not contain extremely high $dE/dx$ events which have very different sources than the backgrounds populating the signal region.  

However, this sideband still contains several different background sources whose models are poorly constrained by other data and must be extrapolated into the signal region.  
The backgrounds are classified as $\nu_e$ CC events, $\nu_\mu$ charged current (CC) interactions, and neutral current interactions, including coherent $\pi^0$ production.  
This sideband is divided into three distinct regions in order to determine overall normalizations for three different background sources, using the energy deposition near the vertex and the electron energy.  
The cuts on the shower end transverse position and the fiducial track length in the hadron calorimeter are removed so the distributions of those observables can be fit over their full ranges.  

In order to minimize potential bias due to mismodeling of energy around a neutrino interaction vertex, the measure of energy deposition used to divide up the sidebands into different regions is different from the one used to isolate the signal events.  $dE/dx_{min}$ is defined as the minimum single-plane $dE/dx$ among the second through sixth planes after the start of the electron candidate track.  The first sideband region contains events with $dE/dx_{min}$ above \unit[3]{MeV/$1.7$cm}.  Because this sideband tends to have more neutral pions, it  has roughly half its events from $\nu_\mu$ CC events, and a third of its events are NC events, with only one sixth expected from from $\nu_e$ events.  
The other two regions have $dE/dx_{min}$ below \unit[3]{MeV/$1.7$cm} but are differentiated by having an electron energy above or below \unit[1.2]{GeV}.  The region with low energy electron candidates is contaminated by $\nu_\mu$ CC events.  With almost three quarters  $\nu_\mu$ CC events, this sideband has only a few per cent $\nu_e$ and one quarter $\nu_\mu$ NC and NC coherent $\pi^0$ production.  The third region, which has low $dE/dx_{min}$ but high electron energy, is about half $\nu_e$ events, with the remainder split between $\nu_\mu$ CC and NC events. In the $\nu_e$-enhanced third region the maximum transverse RMS among the three views is also included in the fit for additional sensitivity to electrons.  

The power of this procedure comes from the fact that the different backgrounds occur in substantially different fractions in each of the three regions.  Because no region of the sideband contains an appreciable fraction of NC coherent $\pi^0$ events, the simulation's prediction for this background cannot be constrained; it is subtracted without modification.  

A $\chi^2$ is formed over all of the distributions and is minimized, allowing three overall background normalizations to float.  The fit returns normalization constants of $0.87\pm 0.03$ for the $\nu_e$ CC backgrounds and $0.58\pm 0.03$ ($0.97\pm 0.02$) for the neutral ( $\nu_\mu$ charged) current backgrounds.  After the fit there is good agreement between the data and simulation for all the distributions used in the fit.  In addition, both the $dE/dx_{min}$ and $E_e\theta^2$ distributions are well-reproduced in the sideband regions after fitting.  


\section{Results}
\label{sec:results}
 
\begin{figure}[tp]
\centering
\ifnum\PRLsupp=0
  \includegraphics[width=\columnwidth]{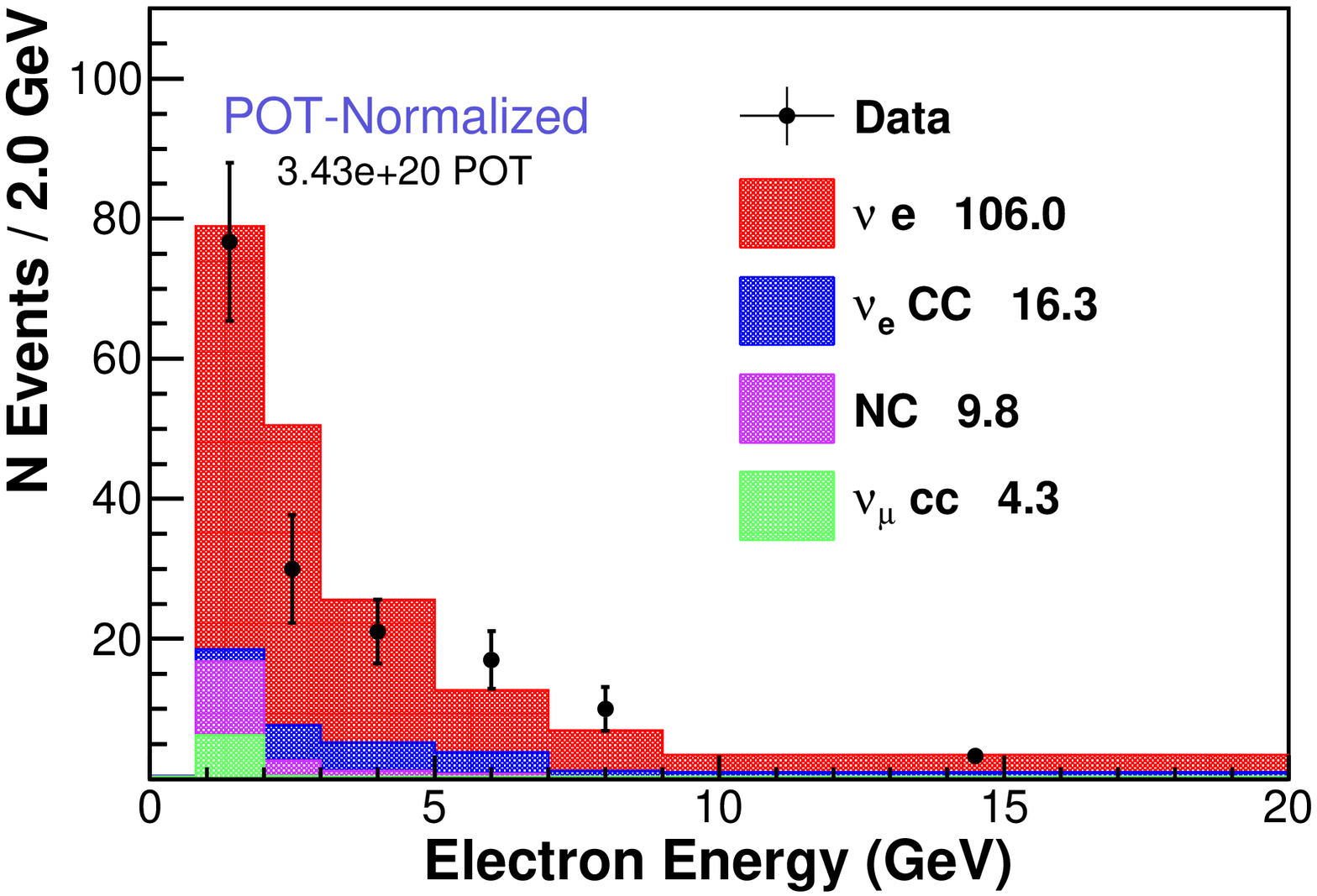}
\else
  \includegraphics[width=0.8\columnwidth]{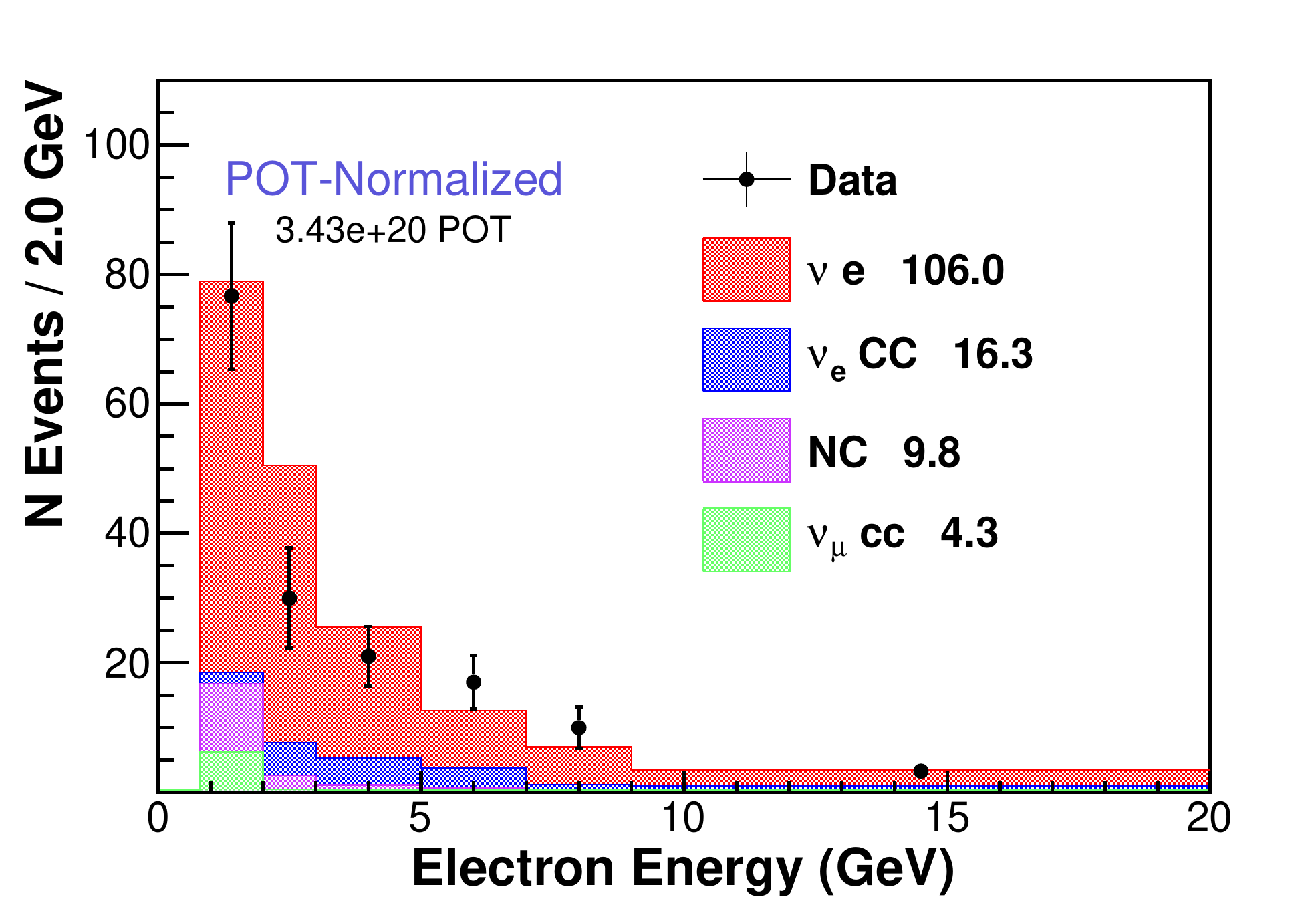} 
\fi
\caption{The electron energy distribution for the data (black points) and predicted signal and backgrounds  (stacked histograms) after all the cuts described in the text are made, and after the background tuning procedure is complete.  Radiative corrections to the $\nu e \rightarrow \nu e$  prediction (described in the Appendix) have been applied. The simulation has been divided according to channel into $\nu_\mu$ and $\nu_e$ scattering on electrons ($\nu$ e), $\nu_e$ charged current interactions ($\nu_e$ CC), other neutral current interactions (NC), and $\nu_\mu$ charged current interactions ($\nu_\mu$ CC).  The numbers in the legend denote the total number of simulated events in each channel after background tuning and radiative corrections.}  
\label{fig:ele_eng}
\end{figure}

\begin{figure}[tp]
\centering
\ifnum\PRLsupp=0
  \includegraphics[width=\columnwidth]{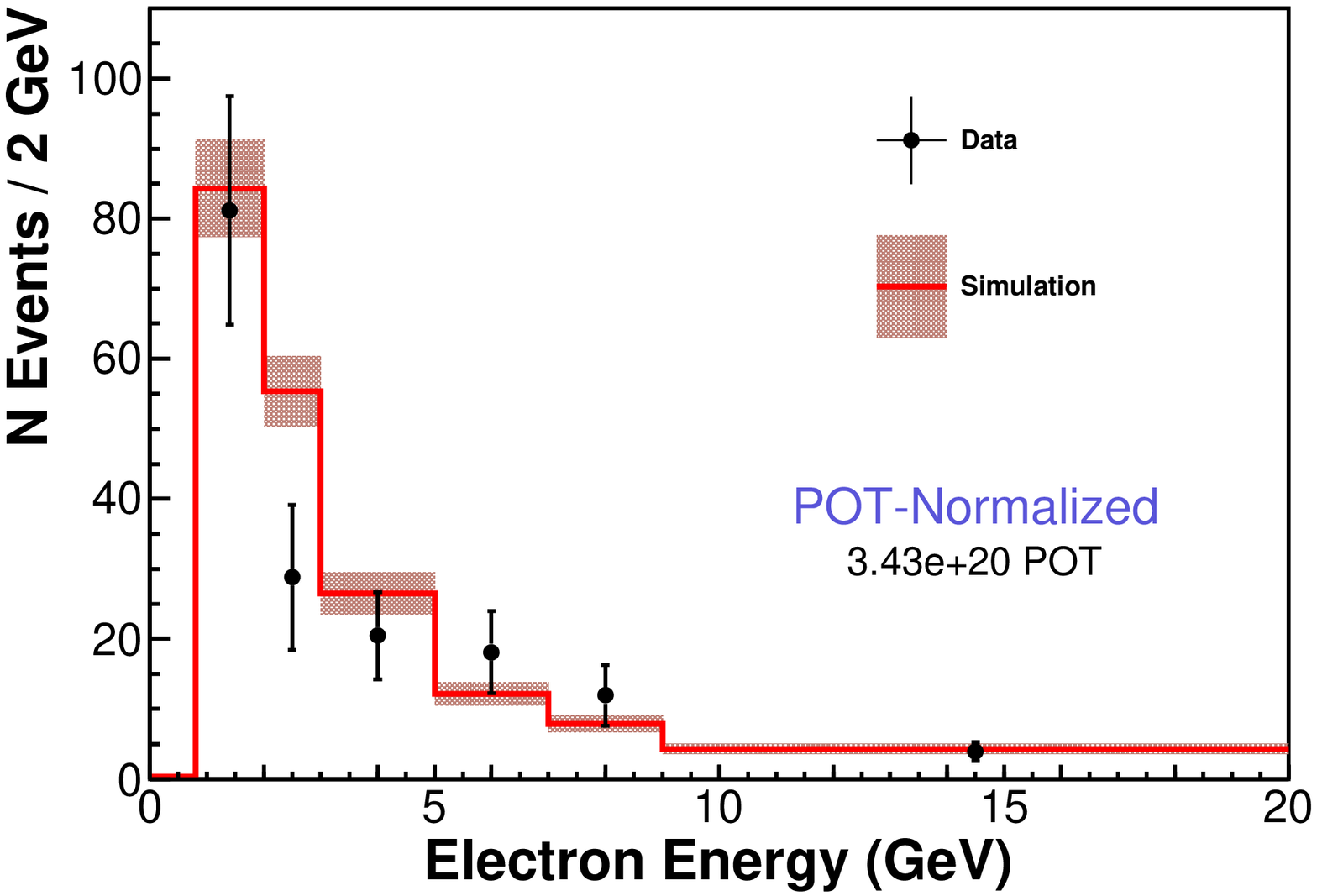}
\else
  \includegraphics[width=0.8\columnwidth]{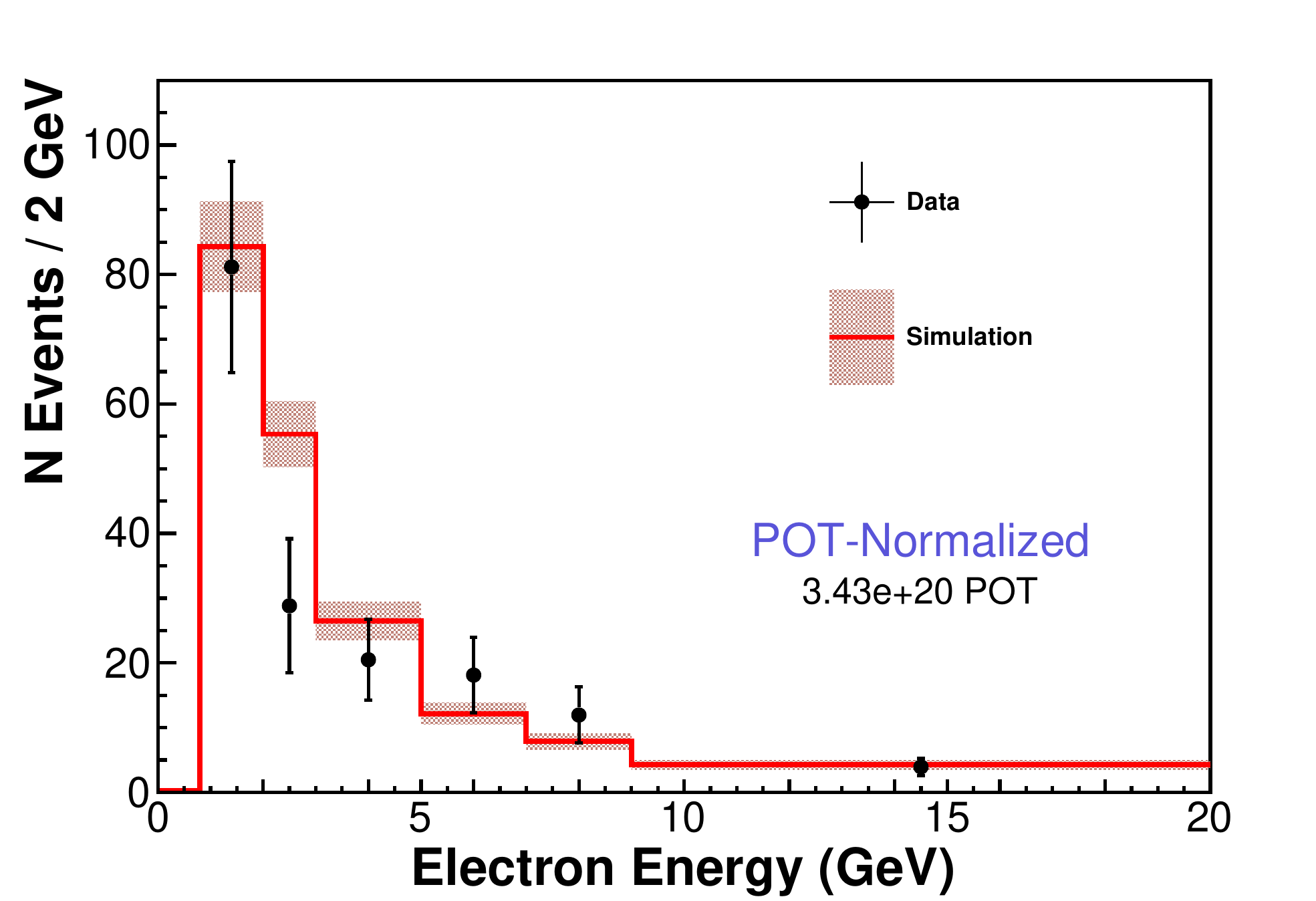} 
\fi
\caption{The electron energy distribution for the data (black) and simulation (red) after all backgrounds are subtracted and after efficiency correction.  Radiative corrections to the $\nu e \rightarrow \nu e$  prediction (described in the appendix) have been applied to the simulation.}
\label{fig:ele_eng_bkg_sub_eff_cor}
\end{figure}

After all the cuts are made, there are a total of 127 candidates, with $30.4\pm 2.3 (stat) \pm 3.3 (syst)$ predicted background events.  The resulting electron energy spectrum is shown in Fig.~\ref{fig:ele_eng}.   
The simulation indicates that the product of acceptance and efficiency averaged across electron energy is $73.3\pm0.5$\% and varies between approximately 70\% at the lower and upper ends of the electron energy spectrum and 78\% at moderate electron energies.  The electron energy spectrum after correction for acceptance and efficiency is shown in Fig.~\ref{fig:ele_eng_bkg_sub_eff_cor}.  The total number of background-subtracted, efficiency-corrected events is $135.3\pm17$.   

\section{Systematic Uncertainties}
\label{sec:systematics}


The total number of neutrino-electron scatters and to a lesser extent the energy distribution of the electrons provide a constraint on the incoming total neutrino flux.  This section describes the uncertainties associated with both the total rate and the spectrum.  
 
\begin{figure}[tp]
\centering
\ifnum\PRLsupp=0
 \includegraphics[width=\columnwidth]{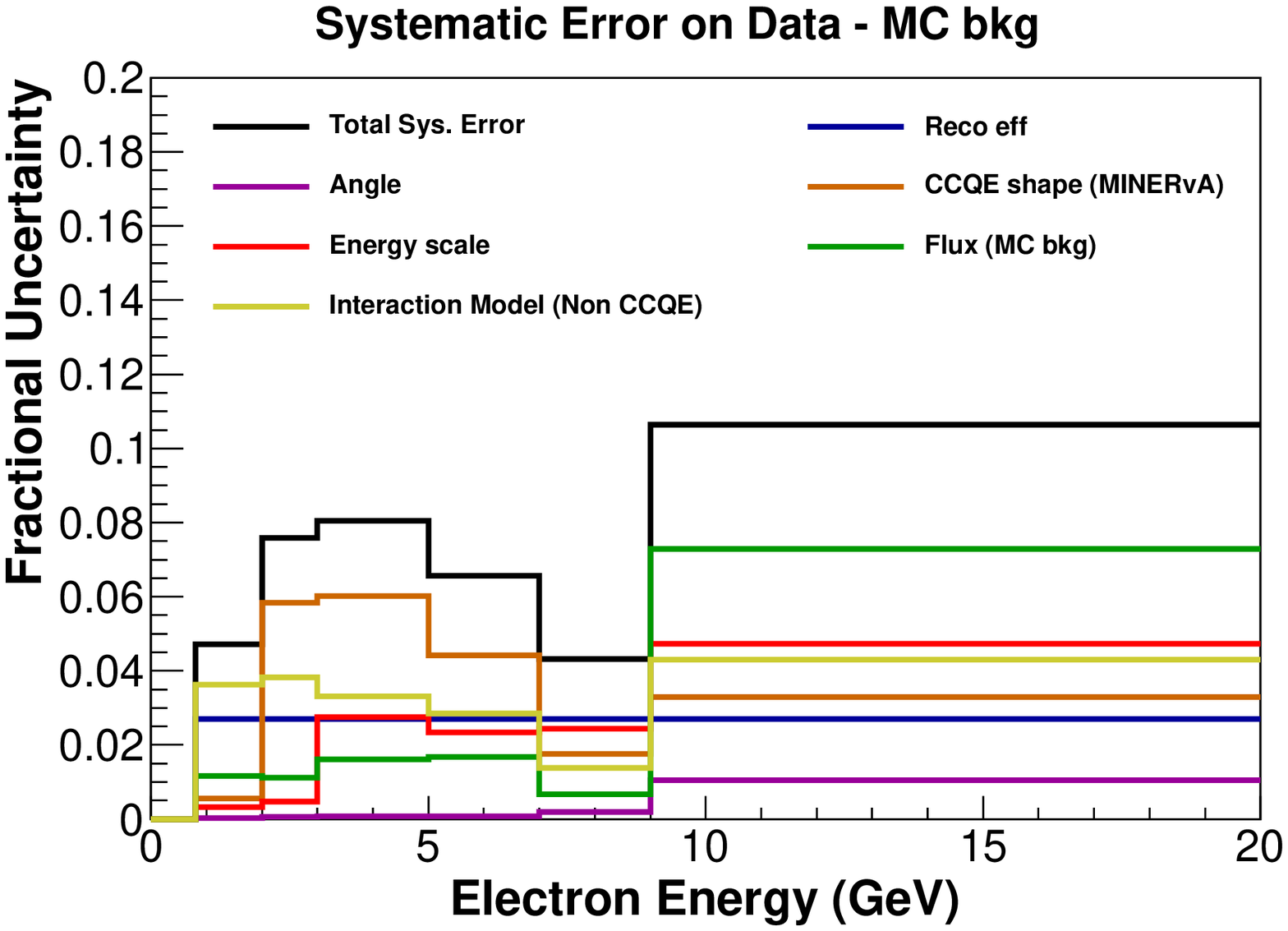}
\else
 \includegraphics[width=0.8\columnwidth]{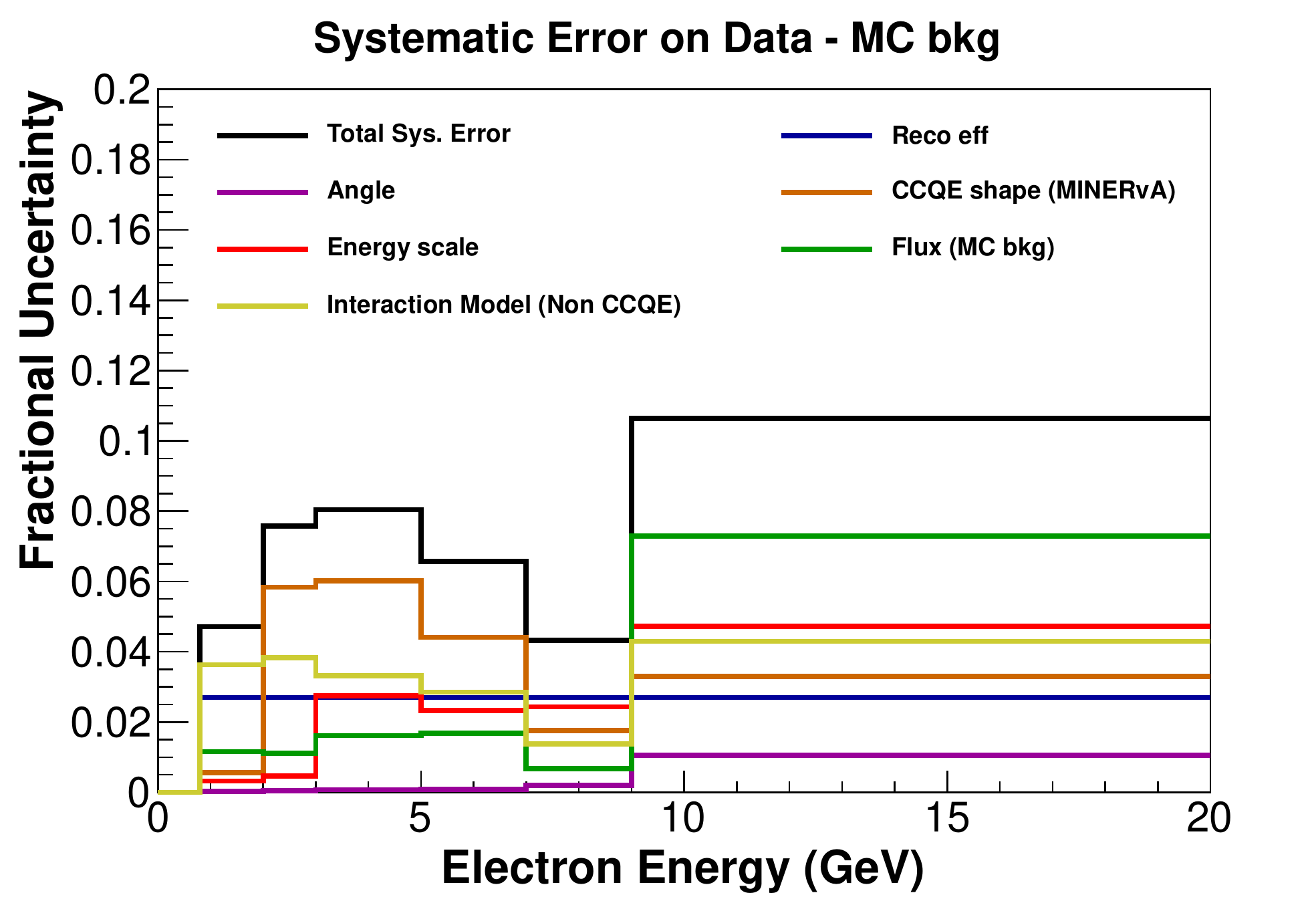}
\fi
\caption{The fractional systematic uncertainties as a function of the electron energy after all the cuts described above are made, and after the tuned background has been subtracted.}
\label{fig:ele_syst}
\end{figure}

The systematic uncertainties can be classified as either the uncertainties in the background prediction or the uncertainties in the detector efficiency and acceptance.  Contributions to the systematic uncertainty in the backgrounds as a function of electron energy are shown in Fig.~\ref{fig:ele_syst}; they are evaluated by changing the underlying simulation prediction according to the various uncertainties, refitting the background scale factors, and then subtracting the background, extracting the electron energy spectrum, and correcting for detector acceptance. 

The largest uncertainty in the background prediction comes from the background cross section models, although it is significantly reduced by the sideband tuning procedure described above.  
The dominant systematic uncertainty for electron energies below 7~GeV comes from the fact that the $\nu_e$ CCQE cross-section shape as a function of $Q^2$ is not known precisely, and for those electron energies the background at low $Q^2$ must be extrapolated using events at high $E_e\theta^2$, which are also at high $Q^2$.  \minerva\ measured a different $\nu_\mu$ cross section shape versus $Q^2$ than what is in the standard GENIE neutrino event generator~\cite{nuprl}, and the systematic is evaluated by taking the difference between the shape of the cross section as a function of  $Q^2$ that \minerva\ measured and the one predicted by GENIE.  There is a recent measurement of the $\nu_e$ CCQE cross-section shape~\cite{jeremy_paper} that shows that within one standard deviation, the $\nu_e$ and $\nu_\mu$ cross-section shapes are consistent with each other.   
At higher electron energies, because of the minimum $Q^2$ cut, this uncertainty no longer dominates and the flux and the electron energy scale become the largest uncertainties.  The flux uncertainties, which contribute primarily to the coherent background subtraction, are incorporated by varying the parameters associated with hadron production and beam focusing in the flux model.  The non-CCQE interaction model uncertainties are evaluated by varying the underlying parameters in the cross-section models for processes such as resonance production and coherent scattering.  

The largest uncertainty in the detector efficiency and acceptance comes from the uncertainty in the electron energy scale (2.2\%).  Although the detector energy scale is set in the simulation using muons from upstream neutrino interactions, there are several other measures which can be used to check agreement for electromagnetic showers.  One such measure is the agreement between data and simulation of Michel electrons which occur when muons stop and decay in the detector.  That energy distribution, which peaks at half the muon mass, was originally 4\% discrepant between data and simulation~\cite{minerva_nim}, although the width of the distribution agreed well between data and simulation.  Another measure is the agreement between data and simulation for the reconstructed neutral pion mass, where the neutral pions are produced along with a muon in a $\nu_\mu$ charged current interaction.  The invariant mass distribution was also discrepant between data and simulation by $5.0\pm2.2\%$~\cite{Trung-pi0} with agreement in the width of the distributions.  Finally, measurements of 400~MeV electrons in our test beam detector also indicated a difference in energy scale but no difference in the resolution between the nominal simulation and the data~\cite{testbeam_nim}.  Although the statistical uncertainty is larger for the neutral pion sample than the other samples, the energies of the photons are much closer to the energies of the electrons for this analysis.  We therefore make a 5\% correction to the electromagnetic energy reconstruction in the data, and assign a $2.2\%$ systematic uncertainty on the absolute energy scale.   

The remaining detector-related uncertainties are associated with the electron angle reconstruction and tracking efficiency.  The uncertainty in the neutrino beam angle direction with respect to the detector axis (1~mrad) is evaluated by comparing the data and simulation for high energy $\nu_\mu$ charged current events that have very low hadronic energy.  Based on that comparison, a correction of 3(1)~mrad is made on the angle in the vertical (horizontal) direction.  The reconstruction efficiency uncertainty is estimated by assuming that the uncertainty for electrons is the same as it is for muons, since both particles' tracks are seeded using the same technique.  The reconstruction efficiency uncertainty for muons is determined by comparing the data and simulation for the efficiency of matching a muon track in MINERvA once a track is found in MINOS that extrapolates into MINERvA.  The discrepancy between data and simulation is treated as the systematic uncertainty.  

The systematic and statistical uncertainties on the total number of neutrino electron scatters is summarized in Tab.~\ref{tab:syst}.  

%
\begin{table}
\begin{tabular}{|l|c|}
\hline 
            &  Fractional  \\   
Source & Uncertainty  \\ \hline
Flux (simulated background)  & 0.2\% \\  \hline
GENIE (not including CCQE) & 2.3\% \\  \hline 
CCQE shape  & 3.1\% \\  \hline
Beam angle & 0.2\% \\  \hline 
Electromagnetic energy scale & 1.8\% \\  \hline 
Reconstruction Efficiency  & 2.7\% \\  \hline 
Total Systematic Uncertainty & 5.1\% \\  \hline \hline 
Statistical Uncertainty  &  12.2\% \\   \hline
\end{tabular}
\caption{Uncertainties associated with the number of events expected after correcting for efficiency.  Sources are described in the text.}
\label{tab:syst}
\end{table}

\section{Flux Constraint}
\label{sec:flux}

Since the total number of neutrino-electron scattering events measured and corrected for efficiency is simply the product of the neutrino-electron scattering cross section, detector mass, and flux, the total uncertainty on the number of signal events collected (shown in Tab.~\ref{tab:syst}) can be thought of as one measurement of an energy-weighted flux integral.  This measurement can be compared to a prediction of that same quantity, where the uncertainties on that prediction include the {\em a priori} flux uncertainties, as well as those coming from the imperfect knowledge of the total number of electrons in the fiducial region (1.5\%) and the uncertainty in the signal cross sections for the different neutrino species in the beam.  

Since the cross sections for electron- and muon-neutrinos differ, as do the cross sections for neutrinos and antineutrinos, as described in the introduction, the cross section used must be an average that is weighted by the relative fractions of all the neutrino species expected in the beam.  The cross sections themselves are known to much better than a tenth of a per cent, and the ratio of electron to muon neutrinos is also well-constrained because most of the electron neutrinos originate from the $\pi^+ \to \mu^+\nu_\mu, \mu^+ \to e^+ \nu_e\bar\nu_\mu$ decay chain.  Therefore, the dominant uncertainty in the {\em a priori} prediction of neutrino-electron scattering events comes from the uncertainty on the flux itself.  With this measurement we are in a regime where the total uncertainty on the measured number of neutrino electron scattering events is comparable to that of the {\em a priori} prediction, so we can use the former in combination with the latter to obtain the most accurate flux prediction.     

In order to incorporate this measurement with the {\em a priori} flux uncertainty quantitatively we make use of Bayes' theorem.  
Following the notation of~\cite{Agashe:2014kda}, Bayes' theorem relates the probability of a hypothesis ($H$) given a data sample ($\mathbf x$) to the product of the probabilty of the hypothesis prior to the measurement ($\pi(H)$) and the probability of the data given the hypothesis ($P({\mathbf x}|H)$):
\begin{equation}
P\left(H | {\mathbf x}\right) = \frac{\pi\left(H\right) P\left({\mathbf x} | H\right) }{\int \pi\left(H'\right) P\left({\mathbf x}|H'\right)dH'},
\end{equation}
where the denominator is a normalization factor.

To use Bayes' theorem to produce a constrained flux prediction, the flux model ($M$) described above is substituted for the hypthesis $H$ and the observed 
number of neutrino-electron scatters is substituted for ${\mathbf x}$, so that:  
\begin{equation}
\label{eq:bayes2}
P(M | N_{\nu e\rightarrow\nu e}) \propto \pi(M) P(N_{\nu e\rightarrow\nu e} | M).
\end{equation}
Thus, the probability of a flux model given the observed absolute electron energy spectrum is proportional to the {\em a priori} probability of that model and the probability of the electron energy spectrum given the model.  The paragraphs below describe how the latter two quantities are computed and combined to form a constraint on the neutrino flux.    

The probability of the neutrino-electron scattering measurement given a model can be estimated by computing a likelihood that assumes the errors on the data in each bin are gaussian-distributed.  This is a good approximation when the number of events in each bin is greater than five, which is the case here.  That likelihood can be expressed as:  
\begin{equation}
\label{eq:gaus_likelihood}
P(N_{\nu e \rightarrow\nu e} | M) = \frac{1}{(2\pi)^{K/2}}\frac{1}{|\Sigma_{\mathbf N}|^{1/2}}e^{-\frac{1}{2}\left({\mathbf N}-{\mathbf M}\right)^T\Sigma_{\mathbf N}^{-1}\left({\mathbf N}-{\mathbf M}\right)}
\end{equation}
~\cite{Lyons86statisticsfor}, where $K$ is the number of bins in the electron energy spectrum, $\mathbf N$ ($\mathbf M$) is the vector representing the bin contents of that spectrum in data (predicted by model M), $\Sigma_N$ is the total data covariance matrix describing all uncertainties on ${\mathbf N}$ except those due to the flux model (available in Table~\ref{tab:result}), and $|\Sigma_N|$ is the determinant of the total covariance matrix.  
 
\begin{figure}[tp]
\centering
\ifnum\PRLsupp=0
  \includegraphics[width=\columnwidth]{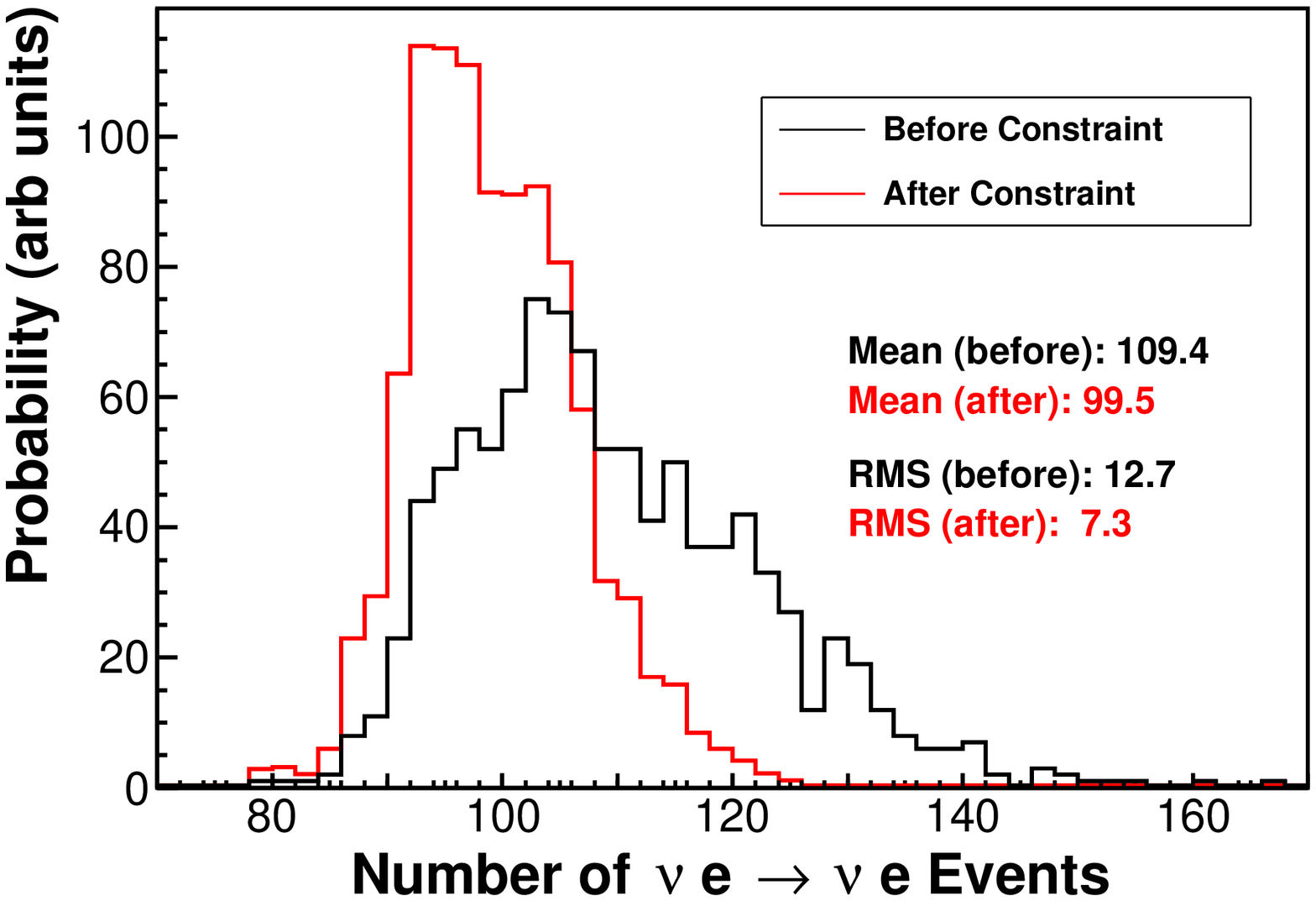}
\else
  \includegraphics[width=0.8\columnwidth]{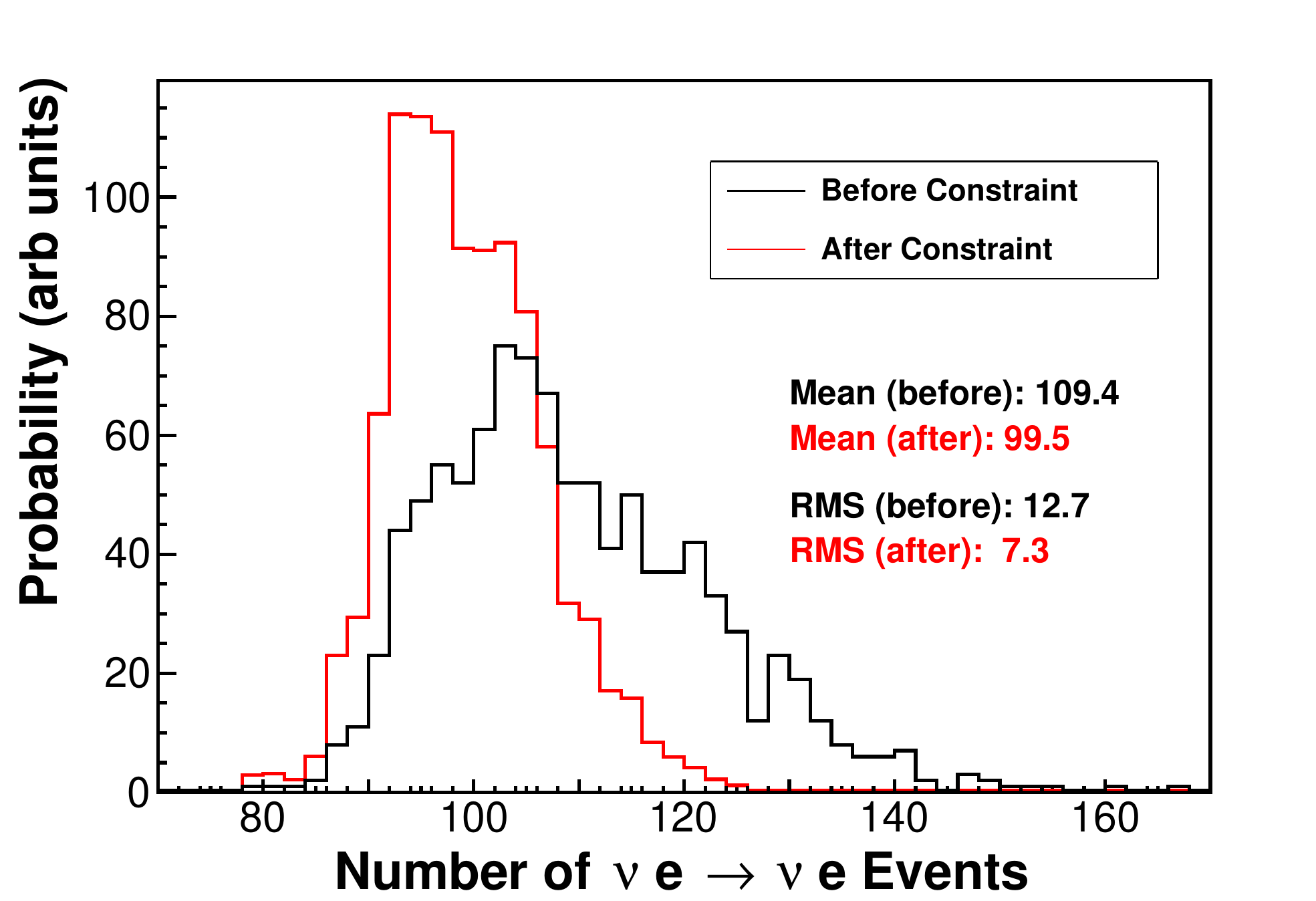}
\fi
\caption{The probability distribution (black) of the predicted total number of neutrino-electron scattering events in the simulation given errors in the neutrino flux model and the modified probability distribution (red) given the observed electron energy spectrum.}
\label{fig:flux_tuning}
\end{figure}

The {\em a priori} (or ``before constraint'') probability distribution of the predicted number of neutrino electron scatters in the \minerva detector is shown in Figure~\ref{fig:flux_tuning}.  It is obtained by randomly varying parameters of the flux simulation within uncertainties repeatedly to produce many ``universes'', each with a different predicted number of neutrino-electron scatters.  The uncertainties in the flux simulation come from external hadron production measurements, uncertainties in the beamline focusing system~\cite{Pavlovic:2008zz}, and comparisons between different hadron production models in regions not covered by external data.  

The constrained probability distribution for the modeled number of neutrino-electron scatters (also shown in Figure~\ref{fig:flux_tuning}) is produced using Equation~\ref{eq:bayes2}.  Specifically, each entry in the {\em a priori} distibution is multiplied by a weight equal to $P(N_{\nu e \rightarrow\nu e} | M)$, evaluated using Equation~\ref{eq:gaus_likelihood}.  The resulting distribution is renormalized to preserve the number of entries in the {\em a priori} distribution.  The constrained number of neutrino-electron scattering events predicted by the model (the mean of the resulting distribution) is lower than the {\em a priori} prediction by 9\%, while the RMS of the constrained distribution is lower by 40\%.

\begin{figure}[tp]
\centering
\ifnum\PRLsupp=0
  \includegraphics[width=\columnwidth]{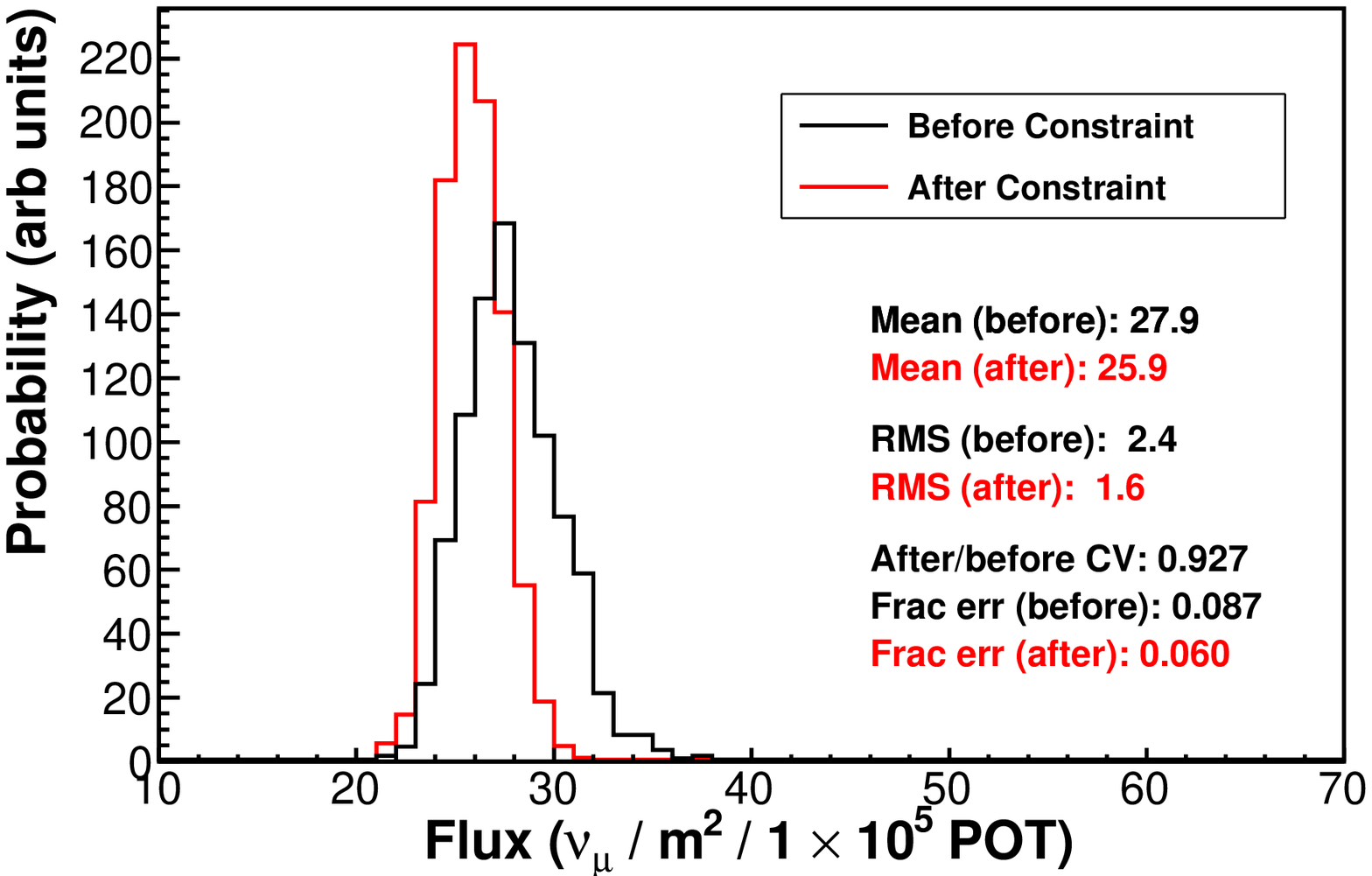}
\else
  \includegraphics[width=0.8\columnwidth]{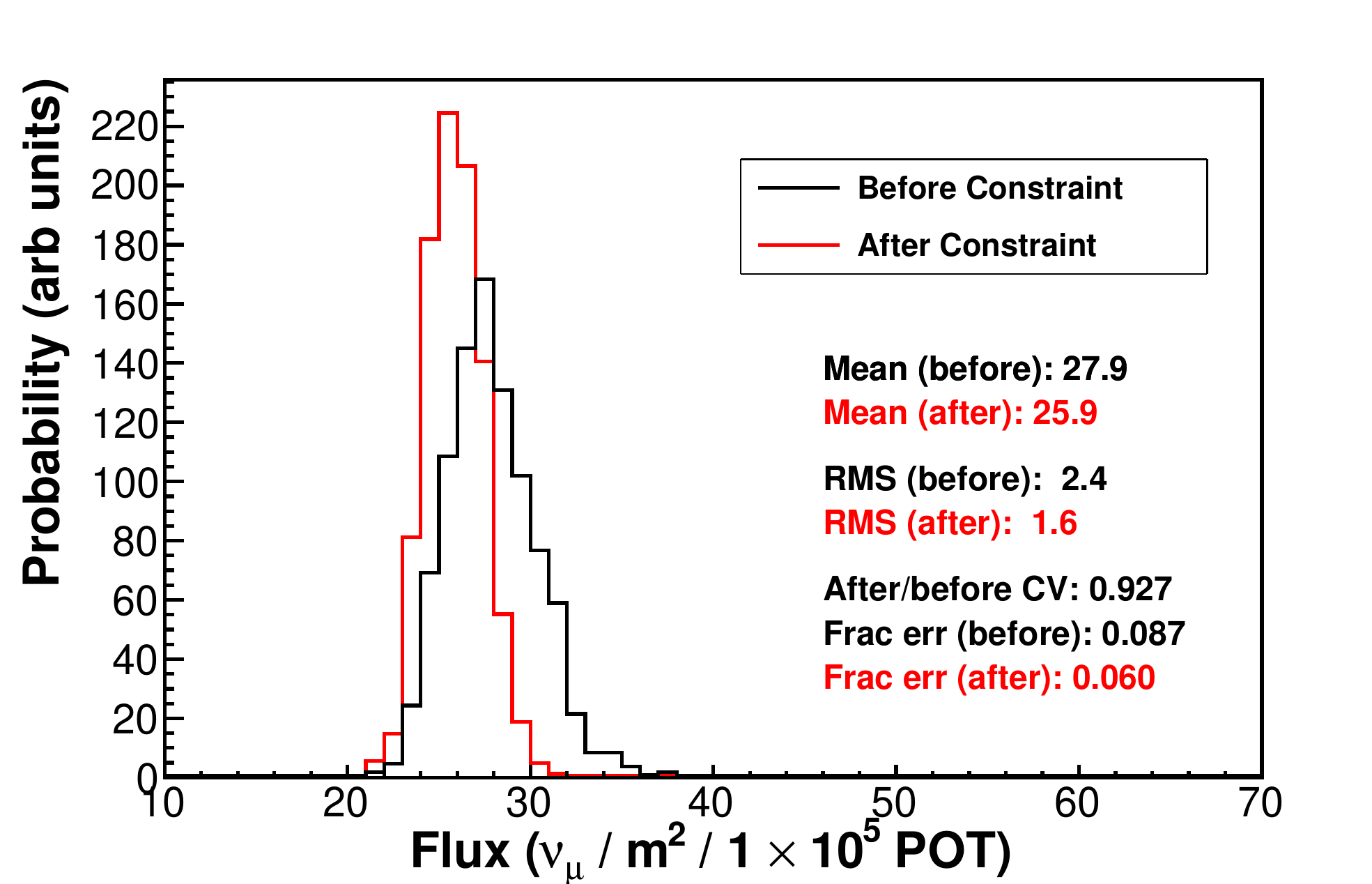}
\fi
\caption{The probability distribution (black) of the predicted $\nu_\mu$ flux integrated between 2 and 10 GeV given errors in the neutrino flux model and the modified probability distribution (red) given the observed electron energy spectrum.}
\label{fig:flux_tuning2}
\end{figure}

The description above uses the predicted number of neutrino-electron scattering events as an example, but the same procedure can be used to constrain any other quantity that is calculable by the simulation and varies depending on the flux prediction.  The {\em a priori} distribution will be different for different quantities, but the weights evaluated using Equation~\ref{eq:gaus_likelihood} are the same regardless of the distribution in question.  For example, the probability distributions of the predicted $\nu_\mu$ flux integrated between 2 and 10 GeV before and after the constraint are shown in Fig.~\ref{fig:flux_tuning2}.  The mean of the constrained probability distribution is lower by 7\% compared to the distribution before the neutrino-electron scattering constraints.  

\begin{figure}[tp]
\centering
\ifnum\PRLsupp=0
\includegraphics[width=\columnwidth]{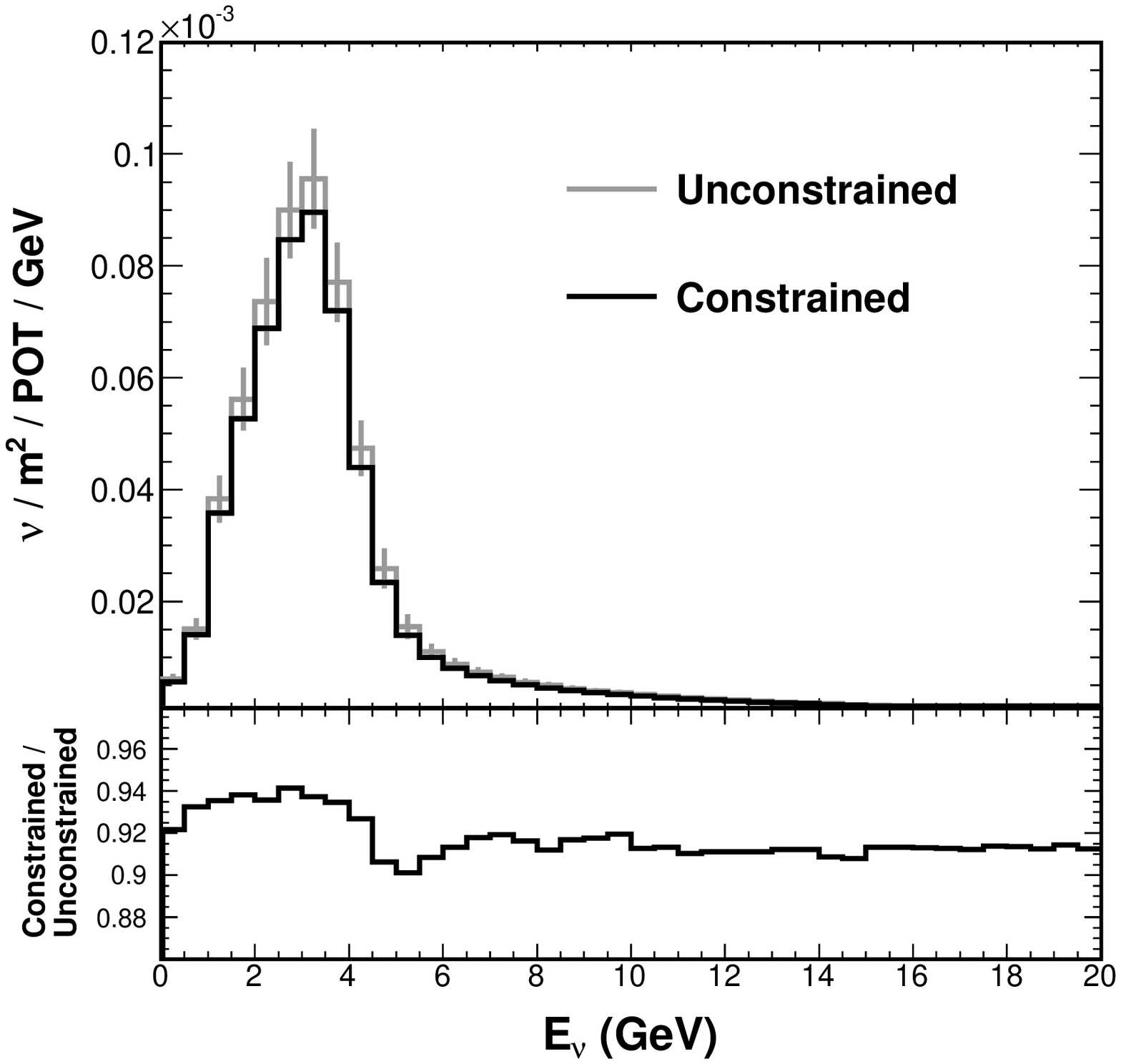}
\else
\includegraphics[width=0.8\columnwidth]{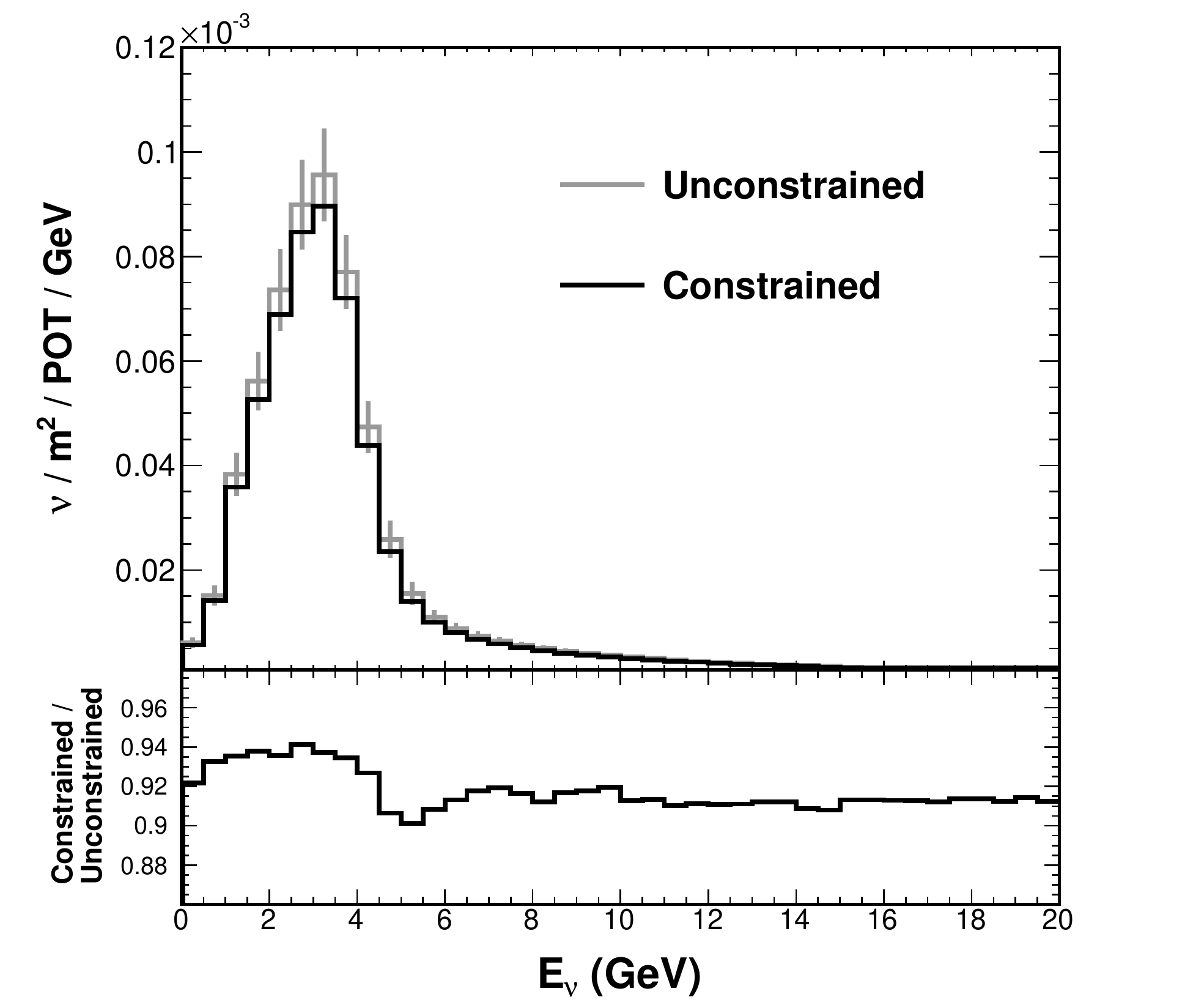}
\fi
\caption{The $\nu_\mu$ flux prediction for the NuMI beamline, before (gray) and after (black) the neutrino-electron scattering flux constraint (top) and the ratio of the constrained to unconstrained predictions (bottom).}
\label{fig:flux_result}
\end{figure}

\begin{figure}[tp]
\centering
\ifnum\PRLsupp=0
\includegraphics[width=\columnwidth]{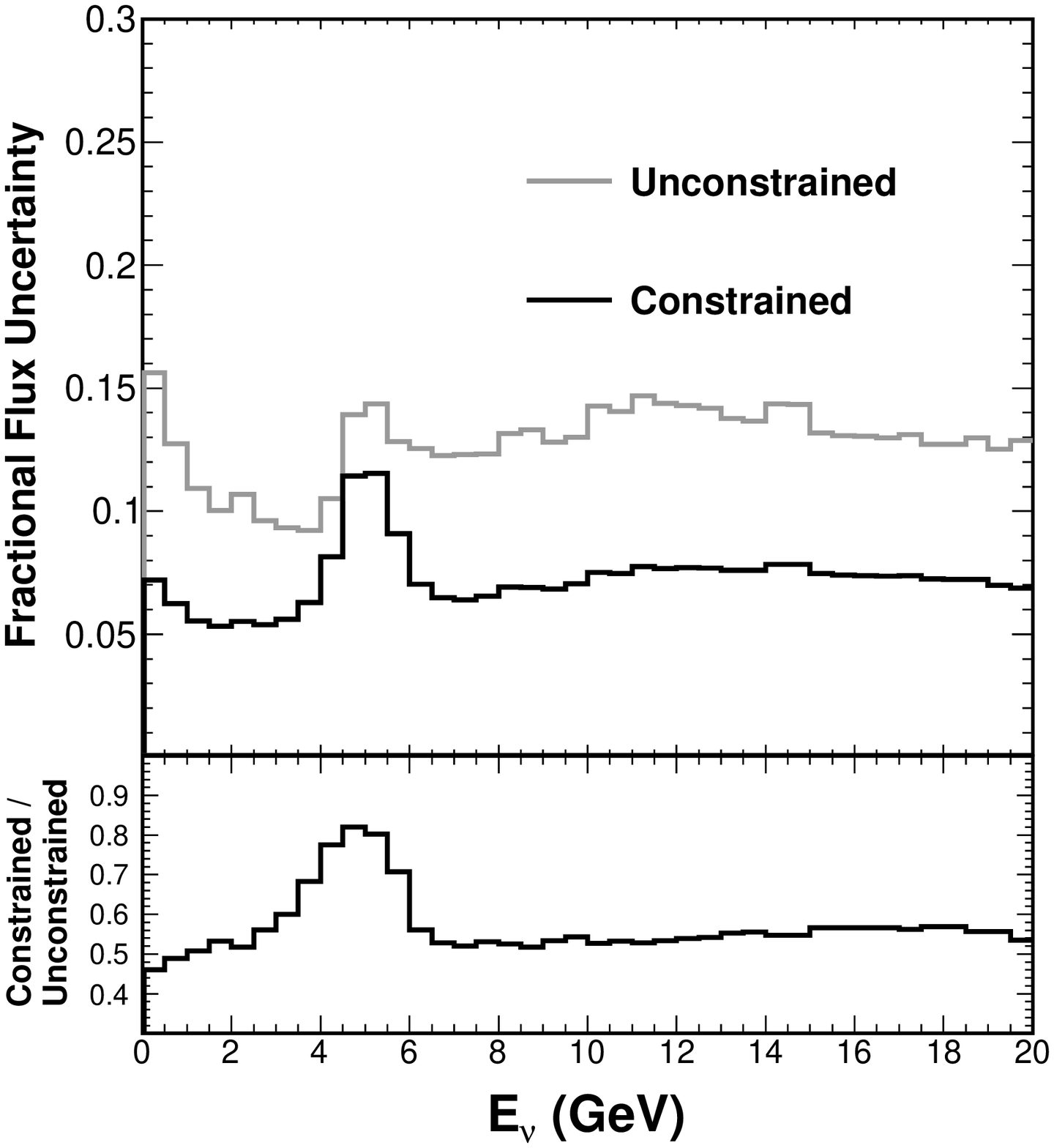}
\else
\includegraphics[width=0.8\columnwidth]{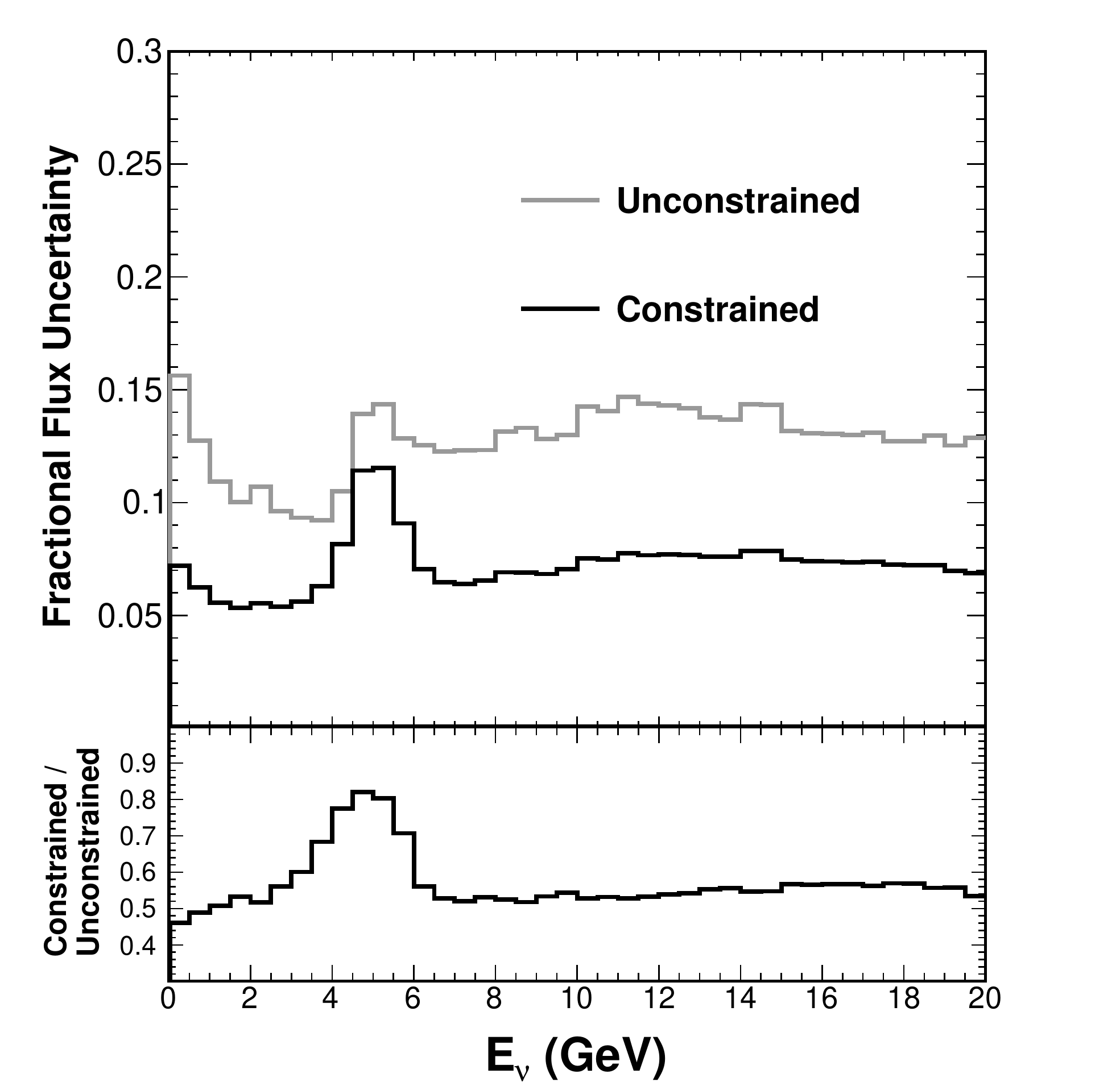}
\fi
\caption{The uncertainty on the $\nu_\mu$ flux prediction for the NuMI beamline, before (gray) and after (black) the neutrino-electron scattering flux constraint (top) and the ratio of the constrained to unconstrained uncertainties (bottom).  The peak at 5~GeV is due to focusing uncertainties which dominate at the falling edge of the neutrino flux, but affect only a small fraction of the neutrino events and therefore are not well-constrained by this technique.}
\label{fig:flux_error}
\end{figure}

The $\nu_{\mu}$ flux as a function of neutrino energy before and after the constraint is shown in Fig.~\ref{fig:flux_result}.  In this case, the procedure described above has been performed separately for each energy bin, and with the constrained flux prediction in each bin taken from the mean of the constrained distributions of fluxes integrated over that energy bin.  The error on this flux, defined as the RMS of the predictions for each neutrino energy bin, before and after the constraint, are shown in Fig.~\ref{fig:flux_error}.  There are large bin-to-bin correlations of the errors on the {\em a priori} flux uncertainty, which are taken into account by the constraint procedure.   Because of the correlations between the $\nu_\mu$ and $\nu_e+\bar\nu_e$ fluxes through the $\pi\rightarrow \nu\mu, \mu\rightarrow \nu_e,\bar\nu_\mu$ chain, this technique can also be used to constrain those fluxes.  For example, MINERvA's measurement of the $\nu_e$ CCQE cross section~\cite{jeremy_paper} uses $\nu_e,\bar\nu_e$ and $\nu_\mu$ flux predictions that have been constrained by the technique described here. 

This procedure assumes that the model and the measurement are compatible.  This can be assessed by evaluating a chisquare between the data and model:
\begin{equation}
\label{eq:chi2}
\chi^2 = \left({\mathbf N}-{\mathbf M}\right)^T\Sigma_{\mathbf N}^{-1}\left({\mathbf N}-{\mathbf M}\right).
\end{equation}
In the case of the data and the model (before constraint) described here, the $\chi^2$ is 9.6 with 6 degrees of freedom, which corresponds to a cumulative probability of 14\%.  This is sufficiently large that the model and data are deemed compatible.

The method described above can be used directly by any other experiment employing the NuMI beam as a neutrino source, regardless of its position or orientation with respect to the beam axis.  To do this, all that is required as input is the predicted electron energy spectrum of neutrino-electron scattering events with $E_{e} >$ \unit[800]{MeV} in a volume and mass corresponding to the MINERvA detector and the same integrated protons on target, distributed according to the assumed uncertainties on the flux prediction.  This should then be compared to the measured efficiency-corrected electron energy spectrum reported by MINERvA.  The MINERvA detector's number of target electrons, location along the NuMI beamline, and volume are given
in Table~\ref{tab:result}.


\begingroup
\squeezetable
\begin{table}
\tabcolsep=0.05cm
\begin{tabular}{c|cccccc}
\hline
$E_e$ (GeV) range& $0.8 - 2$ & $2 - 3$ & $3 - 5$ & $5 - 7$ & $7 - 9$ & $9 - \infty$ \\ 
\hline
$\nu e$ events
& 48.7 & 14.4 & 20.5 & 18.1 & 11.9 & 21.6 \\
in range & $\pm$ 9.9 & $\pm$ 5.2 & $\pm$ 6.3 & $\pm$ 5.9 & $\pm$ 4.3 & $\pm$ 7.7 \\
\hline
$0.8 - 2$    & 98.7  & 1.22  & 1.72  & 1.38  & 0.420 & -0.269 \\
$2 - 3$      & 1.22  & 27.3  & 1.63  & 1.14  & 0.340 & 0.755 \\
$3 - 5$      & 1.72  & 1.63  & 40.1  &  1.88 & 0.596 & 1.35 \\
$5 - 7$      & 1.38  & 1.14  &  1.88 & 34.7  & 0.448 &  0.968 \\
$7 - 9$      & 0.42  & 0.340 & 0.596 & 0.448 & 18.9  & 0.778 \\
$9 - \infty$ &-0.269 & 0.755 &  1.35 & 0.968 & 0.778 & 59.5 \\
\hline
\end{tabular}
\caption{The acceptance-corrected number of $\nu e^-\to\nu e^-$ events in bins of electron energy, their uncertainties, and their covariance matrix.  
The MINERvA detector mass can be represented by a hexagonal prism with face apothem \unit[88.125]{cm} and length \unit[2.53]{m}, oriented with its axis tilted \unit[58]{mrad} upward from the NuMI beam axis, consisting of $1.98 \pm 0.03 \times 10^{30} $ electrons spread uniformly throughout (a fiducial mass of 6.10 tons).  
This volume should be centered at a point \unit[1031.7]{m} from the upstream edge of the first focusing horn in the NuMI beamline and \unit[0.264]{m} (\unit[0.129]{m}) away from the neutrino beam horizontal (vertical) axis in the positive (positive) direction.
}
\label{tab:result}
\end{table}
\endgroup

This measurement is also an important proof-of-principle for a technique that could be used for a future long baseline neutrino experiment, such as the  DUNE~\cite{dune_webpage} experiment.  The process, because it involves scattering off electrons rather than nuclei, provides a precise flux prediction given any near-detector technology.  The only requirements are that the technology provide sufficient angular resolution and energy reconstruction  to isolate these rare events, and that the detector itself have enough fiducial mass to accumulate a statistically significant sample.

\ifnum\sizecheck=0
\begin{acknowledgments}


This work was supported by the Fermi National Accelerator Laboratory
under US Department of Energy contract
No. DE-AC02-07CH11359 which included the \minerva construction project.
Construction support also
was granted by the United States National Science Foundation under
Award PHY-0619727 and by the University of Rochester. Support for
participating scientists was provided by NSF and DOE (USA) by CAPES
and CNPq (Brazil), by CoNaCyT (Mexico), by CONICYT (Chile), by
CONCYTEC, DGI-PUCP and IDI/IGI-UNI (Peru), by Latin American Center for
Physics (CLAF) and by RAS and the Russian Ministry of Education and Science (Russia).  We
thank the MINOS Collaboration for use of its
near detector data. Finally, we thank 
Fermilab for support of the beamline and the detector, and in particular the Scientific Computing Division and the Particle Physics Division for support of data processing.  \end{acknowledgments}
\fi

\bigskip

\ifnum\sizecheck=0
\bibliographystyle{apsrev4-1}
\bibliography{Neutrino-Electron}
\fi

\clearpage
\newcommand{\qsq}{\ensuremath{Q^2_{QE}}\xspace}
\renewcommand{\textfraction}{0.05}
\renewcommand{\topfraction}{0.95}
\renewcommand{\bottomfraction}{0.95}
\renewcommand{\floatpagefraction}{0.95}
\renewcommand{\dblfloatpagefraction}{0.95}
\renewcommand{\dbltopfraction}{0.95}
\setcounter{totalnumber}{5}
\setcounter{bottomnumber}{3}
\setcounter{topnumber}{3}
\setcounter{dbltopnumber}{3}

\appendix{Appendix: One Loop Electroweak Radiative Corrections to Neutrino-Electron Scattering}

The cross-section for tree-level neutrino-electron scattering is given
in Eqn.~\ref{eqn:tree-xsec}, and this is the cross-section implemented
in the GENIE 2.6.2 event generator~\cite{Andreopoulos201087} used as
the reference model in this analysis.  As previously noted, it is necessary to correct this model to use modern values of the electroweak couplings.  This is done by changing the chiral couplings, $C_{LL}$ and $C_{LR}$, to one-loop values predicted using global fits to electroweak data~\cite{Erler:2013xha}.  Table~\ref{tab:ewk-couplings} compares the values for these couplings in GENIE 2.6.2 to the values used in this analysis.

\begin{table}[b]
\begin{tabular}{c|ccc}
   & $C_{\text{LL}}^{\nu_ee}$ & $C_{\text{LL}}^{\nu_\mu e}$ & $C_{\text{LR}}^{\nu e}$ \\ \hline
 \text{GENIE 2.6.2} & 0.7277 & -0.2723 & 0.2277 \\
 \text{One loop} & 0.7276 & -0.2730 & 0.2334 \\
\end{tabular}
\caption{Electroweak couplings in GENIE and in our one-loop calculation of $\nu e^-$ elastic scattering}
\label{tab:ewk-couplings}
\end{table}

In addition, one-loop electroweak radiative corrections ~\cite{Sarantakos:1982bp,Bahcall:1995mm} modify the experessions for the $\nu_\mu e$, $\nubar_\mu e$, $\nu_e e$ and $\nubar_e e$ elastic scattering cross-sections in Eqn.~\ref{eqn:tree-xsec} as follows:
\begin{widetext}
\begin{eqnarray}
\frac{d\sigma(\nu_\ell e^-\to \nu_\ell e^-)}{dy}=&\frac{G_F^2 s}{\pi}&\left[
   \left( C_{LL}^{\nu_\ell e}\right) ^2 (1+\frac{\alpha_{EM}}{\pi}X_1) +\left( C_{LR}^{\nu e}\right) ^2 (1-y)^2 (1+\frac{\alpha_{EM}}{\pi}X_2) \right. \nonumber \\
&&\left. -\frac{C_{LL}^{\nu_\ell e} C_{LR}^{\nu_ e} m y}{E_\nu}(1+\frac{\alpha_{EM}}{\pi}X_3)\right]\\ 
\frac{d\sigma(\nubar_\ell e^-\to \nubar_\ell e^-)}{dy}=&\frac{G_F^2 s}{\pi}&\left[
   \left( C_{LR}^{\nu e}\right) ^2 (1+\frac{\alpha_{EM}}{\pi}X_1) +\left( C_{LL}^{\nu_\ell e}\right) ^2 (1-y)^2 (1+\frac{\alpha_{EM}}{\pi}X_2)  \right. \nonumber \\
&&\left. -\frac{C_{LL}^{\nu_\ell e} C_{LR}^{\nu e} m y}{E_\nu}(1+\frac{\alpha_{EM}}{\pi}X_3)\right]
\end{eqnarray}
\end{widetext}

\noindent  where $E_\nu$ is the neutrino energy, $s$ is the Mandelstam
invariant representing the square of the total energy in the center-of-mass frame, $m$ is the electron mass and $y=T_e/E_\nu$.  The $X_i$ correction terms are

\begin{widetext}
\begin{eqnarray}
X_1&=&\frac{1}{12} (6 y+12 \log (1-y)-6 \log (y)-5) \log \left(\frac{2 E_{\nu }}{m}\right) -\frac{\text{Li}_2(y)}{2}+\frac{y^2}{24}-\frac{11 y}{12} \nonumber \\
&&-\frac{1}{2} \log
   ^2\left(\frac{1}{y}-1\right)+y \log (y) -\frac{1}{12} (6 y+23) \log (1-y)+\frac{\pi ^2}{12}-\frac{47}{36}\\
\nonumber \\
X_2&=&\frac{\left(-4 y^2+\left(-6 y^2+6 y-3\right) \log (y)+11 y+6 (1-y)^2 \log (1-y)-7\right) \log \left(\frac{2 E_{\nu }}{m}\right)}{6
   (1-y)^2} \nonumber \\
&&+\frac{\left(-y^2+y-\frac{1}{2}\right) \left(\text{Li}_2(y)+\log ^2(y)-\frac{\pi ^2}{6}\right)}{(1-y)^2}+\frac{\left(4 y^2+2 y-3\right) \log (y)}{4
   (1-y)^2} \nonumber \\
&&-\frac{31-49 y}{72 (1-y)}+\frac{(10 y-7) \log (1-y)}{6 (1-y)}+\log (1-y) \left(\log (y)-\frac{1}{2} \log (1-y)\right)\\
\nonumber \\
X_3&=&\log \left(-\frac{m}{\sqrt{y E_{\nu } \left(2 m+y E_{\nu }\right)}+m+y E_{\nu }}+1-y\right) \nonumber  \\
&& \times \left(\frac{\left(m+y E_{\nu }\right) \log \left(\frac{\sqrt{y
   E_{\nu } \left(2 m+y E_{\nu }\right)}+m+y E_{\nu }}{m}\right)}{\sqrt{y E_{\nu } \left(2 m+y E_{\nu }\right)}}-1\right)
\end{eqnarray}
\end{widetext}

\noindent where Li$_2(z)$ represents Spence's function, $\int_0^z\frac{-\log(1-u)}{u}du$.


\end{document}